\documentclass[aps,prb,amsmath,amssymb,amsfonts]{revtex4}
\usepackage[caption=false]{subfig}
\usepackage{graphicx}
\usepackage{color}
\usepackage{booktabs,array,multirow}
\usepackage[outdir=./]{epstopdf}
\newcommand{\rn}[1]
  {\MakeUppercase{\romannumeral #1}}

\begin{document}

\title{On the Optical Properties of Excitons in Buckled 2D Materials in an External Electric Field}
%\author{Matthew Brunetti}
%\affiliation{City Tech}
%\author{Oleg Berman}
%\affiliation{City Tech, Graduate Center}
%\author{Roman Kezerashvili}
%\affiliation{City Tech, Graduate Center}
\author{Matthew N. Brunetti$^{1,2}$, Oleg L. Berman$^{1,2}$, and Roman Ya. Kezerashvili$^{1,2}$}
\affiliation{%
$^{1}$Physics Department,  New York City College of Technology\\
The City University of New York,
  300 Jay Street,   Brooklyn NY, 11201, USA \\
$^{2}$The Graduate School and University Center\\
The City University of New York,
New York, NY 10016, USA \\
}

\date{\today}

\begin{abstract}
  We study the binding energies and optical properties of direct and indirect excitons in monolayers and double layer heterostructures of Xenes: silicene, germanene, and stanene.
  It is demonstrated that an external electric field can be used to tune the eigenenergies and optical properties of excitons by changing the effective mass of charge carriers.
  The Schr\"{o}dinger equation with field-dependent exciton reduced mass is solved by using the Rytova-Keldysh (RK) potential for direct excitons, while both the RK and Coulomb potentials are used for indirect excitons.
  It is shown that for indirect excitons, the choice of interaction potential can cause huge differences in the eigenenergies at large electric fields and significant differences even at small electric fields.
  Furthermore, our calculations show that the choice of material parameters has a significant effect on the binding energies and optical properties of direct and indirect excitons.
  These calculations contribute to the rapidly growing body of research regarding the excitonic and optical properties of this new class of two dimensional semiconductors.
\end{abstract}

\maketitle

\section{\label{sec:intro}Introduction}

  Following the discovery of stable graphene monolayers in 2004~\cite{novoselov2004electric} and the subsequent isolation of two-dimensional (2D) insulators such as hexagonal boron nitride ($h$-BN)~\cite{Dean2010} and 2D semiconductors such as transition metal dichalcogenides (TMDCs)~\cite{Mak2010}, researchers have continually sought to discover new 2D materials with novel properties.
  One recent addition to the 2D universe are the group \rn{14} elements, namely silicon, germanium, and tin, whose 2D forms are referred to as silicene, germanene, and stanene (or sometimes, tinene), respectively.
  A recent paper~\cite{Molle2017} referred to buckled 2D monolayers consisting of silicon, germanium, and tin as ``Xenes''.
  For the sake of brevity, we shall adopt the same convention when collectively referring to the behavior or properties of silicene, germanene, and stanene.
  Early theoretical studies~\cite{Fagan2000,Cahangirov2009,lebegue2009} were soon followed by the first experimental reports of silicene nanoribbons~\cite{Aufray2010,DePadova2010} and 2D silicene sheets~\cite{Lalmi2010}.

  These early studies of silicene revealed one of the most crucial differences between silicene, the heavier group~\rn{14} elements germanene and stanene, and graphene: silicene's most stable form is not a perfectly flat sheet, but is instead slightly buckled~\cite{Vogt2012,Lin2012,Li2014d}.
  Among other novel phenomena, this buckling allows one to tune the band gap of Xenes by applying an external electric field perpendicular to the plane of the monolayer~\cite{Ni2012,Stille2012,Drummond2012,Fadaie2016,Gao2014c}.
  The tunable band gap of Xenes gives researchers, among other things, extraordinary \textit{in situ} control over the  binding energies and optical properties of excitons in these materials.

  In general, excitons in 2D materials are interesting because of their potential for large binding energies, strong optical absorption, and unique collective properties.
  Indeed, excitons in TMDCs are characterized by relatively high binding energies and significant spin-orbit coupling~\cite{Berkelbach2013,Kormanyos2015}.
  Bose-Einstein condensation and superfluidity of spatially indirect excitons in TMDC/$h$-BN heterostructures, formed by two TMDC monolayers separated by $N$ $h$-BN monolayers, were also analyzed~\cite{Fogler2014,Wu2015,Berman2016,Berman2017a}.
  A theoretical study of intraexcitonic optical transitions in TMDC/$h$-BN heterostructures was performed~\cite{Brunetti2018}.
  Recently, experimental studies of the excited states of direct excitons in monolayer MoS$_2$~\cite{Robert2017} and spatially indirect excitons in multi-layer MoSe$_2$ single crystals~\cite{Horng2018} have also been performed.
  A comprehensive review of excitons in TMDCs is given in Ref.~\onlinecite{Wang2018}.
  While there is an abundance of research regarding excitons in TMDCs, there is relatively little research on excitons in buckled 2D materials.

  Experimental studies of intraexcitonic optical transitions have been performed in Cu$_2$O~\cite{Kuwata-Gonokami2004,Kubouchi2005,Jorger2005,Huber2006}, and GaAs/GaAlAs semiconductor coupled quantum wells~\cite{Huber2005,Huber2008}.
  Recently, similar experiments have been performed on direct excitons in monolayer TMDCs~\cite{Cha2016,Poellmann2015b,Steinleitner2017}, but there are not yet any comparable studies of the 2D Xenes.
  In this paper we perform a theoretical study of the binding energies and optical properties of direct and indirect excitons in buckled 2D materials under the effect of an external electric field.

  The objective of this paper is to study the exceptional tunability, via application of an external electric field, of the properties of excitons in Xenes.
  We demonstrate this by explicitly calculating the binding energies and optical properties of excitons in the case of (i) direct excitons in Xene monolayers and (ii) spatially indirect excitons in heterostructures formed by two Xene monolayers separated by $N$ monolayers of $h$-BN, which we denote as X-BN-X.

  First, in the framework of the effective mass approximation, we consider the dependence of the exciton reduced mass $\mu$ as a function of the perpendicular external electric field, $E_\perp$.
  This field-dependent mass is used when solving the Schr\"{o}dinger equation for the eigenfunctions and eigenenergies of the direct or indirect exciton.
  This allows us to furnish relevant optical quantities such as the oscillator strength and absorption coefficient.
  Second, we investigate the dependence of the binding energies and optical properties of direct excitons in monolayer Xenes on the external electric field.
  For spatially indirect excitons in X-BN-X heterostructures, we study the dependence of these quantities on the interlayer separation as well as on the external electric field.

  This Paper is organized in the following way.
  In Sec.~\ref{sec:theory}, we present a theoretical framework for excitons in buckled 2D materials within the effective mass approach and consider their optical properties when the effective mass is electric-field dependent.
  The binding energies and optical properties of direct excitons in monolayer Xenes and of indirect excitons in X-BN-X heterostructures are presented in Secs.~\ref{sec:dirres} and~\ref{sec:ind}, respectively.
  A comparison between direct and indirect excitons is given in Sec.~\ref{sec:dirindcomp}.
  Our conclusions follow in Sec.~\ref{sec:conc}.

  \section{\label{sec:theory}Theoretical framework of 2D Excitons with electric field-dependent mass}

  \subsection{\label{ssec:theoryhamiltonian}Charge Carriers in buckled 2D materials}

  Monolayers of silicene, germanene, and low-buckled stanene can be pictured as graphene monolayers in which the two triangular sublattices are offset with respect to the plane of the monolayer by a particular distance, known as the buckling constant or buckling factor.
  This offset between the two triangular sublattices gives rise to the intrinsic sensitivity of Xenes to an external electric field applied perpendicular to the plane of the monolayer.
  With no external electric field, the band structure of Xenes in the vicinity of the $K/K'$ points resembles graphene, though the intrinsic gaps of Xenes are significantly larger than that of graphene.
  The application of a perpendicular electric field creates a potential difference between the sublattices, causing a change in the band gap in the material, which in turn changes the effective masses of the electrons and holes.

  The Hamiltonian in the vicinity of the $K/K'$ points is given in Ref.~\onlinecite{Tabert2014} as:
  \begin{equation}
	\hat{H} = \hbar v_F \left( \xi k_x \hat{\tau}_x + k_y \hat{\tau}_y \right) - \xi \Delta_{gap} \hat{\sigma}_z \hat{\tau}_z + \Delta_z \hat{\tau}_z,
	\label{eq:taberthamiltonian}
  \end{equation}
  where $v_F$ is the Fermi velocity, $\hat{\tau}$ and $\hat{\sigma}$ are the pseudospin and real spin Pauli matrices, respectively, $k_x$ and $k_y$ are the components of momentum in the $xy$-plane of the monolayer, relative to the $K$ points, $2\Delta_{gap}$ is the intrinsic band gap, $\xi,\sigma = \pm 1$ are the valley and spin indices, respectively, and $\Delta_z = e d_0 E_\perp$ is the gap induced by the electric field, $E_\perp$, acting along the $z$-axis, where $d_0$ in the latter expression is the buckling constant.
  The first term in Eq.~\eqref{eq:taberthamiltonian} is the same as that of the low-energy Hamiltonian in graphene~\cite{CastroNeto2009a,Abergel2010}.
  The second term describes the spin-orbit coupling~\cite{Kane2005} with an intrinsic band gap of $2\Delta_{gap}$, while the last term describes the modification of the intrinsic band gap via an external electric field.

  Using Eq.~\eqref{eq:taberthamiltonian}, one may write the dispersion relation of charge carriers near the $K/K'$ points as:
  \begin{equation}
	E(k) = \sqrt{\Delta_{\xi \sigma}^{2} + \hbar^2 v_F^2 k^2},
	\label{eq:disprel}
  \end{equation}
  where
  \begin{equation}
	\Delta_{\xi \sigma} = \vert \xi \sigma \Delta_{gap} - e d_0 E_\perp \vert
	\label{eq:deltaez}
  \end{equation}
  is the electric field-dependent band gap at $k=0$.
  We note that when $E_\perp = 0$, the spin-up and spin-down bands of the valence and conduction bands are degenerate.
  In other words, spin-orbit splitting only manifests itself at non-zero external electric fields.
  At non-zero electric fields, both the valence and conduction bands split, into upper bands with a large gap (when $\xi = -\sigma$), and lower bands with a small gap (when $\xi=\sigma$).
  We call the excitons formed by charge carriers from the large gap $A$ excitons, and those formed by charge carriers in the small gap $B$ excitons.
  When the external field reaches a critical value $E_c = \Delta_{gap}/(e d_0)$, the lower bands form a Dirac cone at the $K/K'$ points.
  The values of these quantities are presented in Table~\ref{tab:matpars}.

  \begin{table*}[t]
	\centering
	\begin{tabular}{|r|c|c|c|c|c|}
	  \hline
	  Material									&	$2\Delta_{\text{gap}}$ (meV)&	$d_0$ (\AA)					&	$v_F$ ($\times 10^5~\text{m/s}$)	&	$\epsilon$	&	$l$ [nm]\\\hline
	  Silicene (FS)								&	1.9~\cite{Matthes2013}		&	0.46~\cite{Ni2012}			&	6.5~\cite{Matthes2013}	&	11.9	&	0.4~\cite{Tao2015}\\\hline
	  Silicene ($h$-BN, Type I)~\cite{Li2013}	&	27							&	0.46						&	4.33					& 	11.9	&	0.333\\\hline
	  Silicene ($h$-BN, Type II)~\cite{Li2013}	&	38							&	0.46						&	5.06					& 	11.9	&	0.333\\\hline
	  Germanene (FS)							&	33~\cite{Matthes2013}		&	0.676~\cite{Ni2012}			&	6.2~\cite{Matthes2013}	&	16	&	0.45\\\hline
	  Stanene (FS)								&	101~\cite{Matthes2013}		&	0.85~\cite{Balendhran2015}	&	5.5~\cite{Matthes2013}	&	24	&	0.5\\
	  \hline
	\end{tabular}
	\caption{\label{tab:matpars}%
	  Parameters for buckled 2D materials:
	  $2\Delta_{\text{gap}}$ is the total gap between the conduction and valence bands;
	  $d_0$ is the buckling parameter;
	  $v_F$ is the Fermi velocity;
	  $l$ is the monolayer thickness;
	  $\epsilon$ is the dielectric constant of the bulk material.
	  FS refers the freestanding Xene monolayers.
	}
  \end{table*}

  In the vicinity of the $K/K'$ points, the conduction and valence bands are parabolic.
  The effective mass of charge carriers near the $K/K'$ points can be written as $m^{*} = \Delta_{\xi \sigma}/v_F^2$.
  The effective masses of electrons and holes are the same due to the symmetry between the lowest conduction and highest valence bands, and can be written in terms of the external electric field as:
  \begin{equation}
	m^{*} = \frac{\lvert \xi\sigma\Delta_{gap} - e d_0 E_\perp \rvert}{v_F^2}.
	\label{eq:effmassEz}
  \end{equation}

  \begin{figure*}[h]
	\centering
	\includegraphics[width=1\columnwidth]{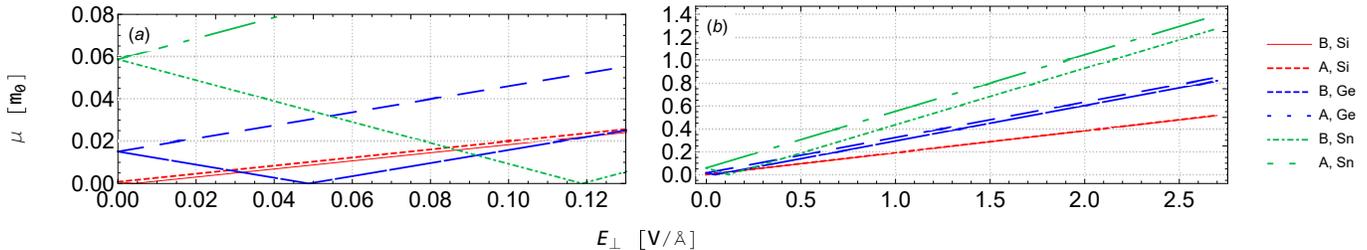}
	\caption{%
	  Exciton reduced mass $\mu$, in units of $m_0$, as a function of the external electric field, $E_\perp$.
	  (a) $\mu$ as a function of $E_\perp$, zoomed in to show the behavior at small values of $E_\perp$.
	  (b) $\mu$ as a function of $E_\perp$ across the full range of $E_\perp$ considered in the calculations.
	}
	\label{fig:muvsez}
  \end{figure*}

  The behavior of $\mu$ as a function of $E_\perp$ for $A$ and $B$ excitons in freestanding (FS) Xenes is shown in Fig.~\ref{fig:muvsez}.
  Following \textit{ab initio} calculations~\cite{Drummond2012} which determined that the crystal structure of silicene becomes unstable around 2.6 V/\AA, we consider in our calculations electric fields up to 2.7 V/\AA.
  As one can see from Table~\ref{tab:matpars}, silicene, which has the largest $v_F$ and the smallest $d_0$, has the smallest slope.
  Even at $E_\perp = 2.7~\text{V/\AA}$, the exciton reduced mass in silicene never surpasses the electron rest mass, $m_0$.
  On the other hand, in stanene, which has the smallest $v_F$, the exciton reduced mass exceeds $m_0$ at large fields.

  At small electric fields, germanene and especially stanene show significant differences between the reduced masses of the $A$ and $B$ excitons \textendash{} this is due to their large intrinsic band gaps.
  Silicene, which has an intrinsic band gap on the order of a couple meV, exhibits very little difference between the reduced masses of $A$ and $B$ excitons, even at relatively small electric fields.
  At large electric fields the difference between the $A$ and $B$ exciton reduced mass is negligible in silicene and germanene.
  In all cases, the mass of the $A$ exciton exceeds the mass of the $B$ exciton.

  \subsection{\label{ssec:theoryexcitons}Effective mass approach for excitons in buckled 2D materials}

  In order to obtain the eigenfunctions and eigenenergies of an exciton in Xenes, we first write the Schr\"{o}dinger equation for an interacting electron and hole:
  \begin{equation}
	\left[ \frac{- \hbar^2}{2 m_e} \nabla^2_e + \frac{- \hbar^2}{2 m_h} \nabla^2_h + V\left( r_e, r_h \right) \right] \psi \left( r_e, r_h \right) = E \psi \left( r_e, r_h \right),
	\label{eq:ehschro}
  \end{equation}
  where $e$ and $h$ are the indices referring to the electron and hole, respectively, $m_e = m_h = m^{*}$ are the masses of charge carriers given by Eq.~\eqref{eq:effmassEz}.
  Performing the standard procedure for the coordinate transformation to the center-of-mass, $R_{CM} = (m_e r_e + m_h r_h)/(m_e + m_h)$, and relative coordinates, $r = r_e - r_h$, one obtains an equation for the relative motion of the electron and hole:
  \begin{equation}
	\left[ \frac{- \hbar^2}{2 \mu} \nabla^2 + V(r) \right] \psi(r) = E \psi(r),
	\label{eq:relschro}
  \end{equation}
  where $\mu = m_e m_h/(m_e + m_h) = m^{*}/2$ is the exciton reduced mass.

  The relative separation $r$ between the electron and hole can be written in cylindrical coordinates, $r = \rho \hat{\rho} + D \hat{z}$, allowing us to treat the case of direct excitons in an Xene monolayer and spatially indirect excitons in X-BN-X heterostructures on equal footing.
  For direct excitons, we set $D=0$, and Eq.~\eqref{eq:relschro} becomes a purely 2D equation, with $\rho$ representing the separation between the electron and hole sharing the same plane.
  Throughout this paper, we consider the separation between two Xene monolayers in steps of calibrated thickness, $l_{h-\text{BN}} = 0.333~\text{nm}$, corresponding to the thickness of one $h$-BN monolayer.
  For spatially indirect excitons, the relative coordinate $r = \sqrt{\rho^2 + D^2}$, where $D= l + N l_{h-\text{BN}}$, $l$ is the Xene monolayer thickness and $N$ is the number of $h$-BN monolayers.

  The interaction between the electron and hole in 3D homogeneous dielectric environments is described by the Coulomb potential, but this interaction in 2D is modified and described by a potential which includes screening effects as a result of the reduced dimensionality.
  This potential was first considered in Ref.~\onlinecite{Rytova1967} and was independently derived in Ref.~\onlinecite{Keldysh1979}~\textendash{} we refer to it as the Rytova-Keldysh (RK) potential.
  Therefore the interaction potential $V(r)$ between the electron and hole for direct excitons in an Xene monolayer is:
  \begin{equation}
	V(r) = \frac{\pi k e^2}{2 \kappa \rho_0} \left[ H_0 \left( \frac{r}{\rho_0} \right) - Y_0 \left( \frac{r}{\rho_0} \right) \right]
	\label{eq:keldyshpot}
  \end{equation}
  where $\rho_0$ is the screening length, and $H_0$ and $Y_0$ are the Struve and Bessel functions of the second kind, respectively.
  In Eq.~\eqref{eq:keldyshpot}, $\kappa = (\epsilon_1 + \epsilon_2)/2$ describes the surrounding dielectric environment, $\epsilon_1$ and $\epsilon_2$ are the dielectric constants either (a) above and below the monolayer, in the case of direct excitons in an Xene monolayer, or (b) between and surrounding the Xene monolayers in the case of spatially indirect excitons in an X-BN-X heterostructure.
  The screening length $\rho_0$ can be written as~\cite{Berkelbach2013} $\rho_0 = (2 \pi \chi_{2D})/(\kappa)$, where $\chi_{2D}$ is the 2D polarizability, which in turn is given by~\cite{Keldysh1979} $\chi_{2D} = \left( l \epsilon \right)/ (\epsilon_1 + \epsilon_2)$, where $\epsilon$ is the bulk dielectric constant of the Xene monolayer.

  To better understand the importance of the screening effect in X-BN-X heterostructures, we perform calculations using both the RK and Coulomb potentials.
  For indirect excitons, the expressions for the interaction between the electron and hole can be written as:

  \begin{equation}
	V(\sqrt{\rho^2 + D^2}) = \frac{\pi k e^2}{2 \kappa \rho_0} \left[ H_0 \left( \frac{\sqrt{\rho^2 + D^2}}{\rho_0} \right) - Y_0 \left( \frac{\sqrt{\rho^2 + D^2}}{\rho_0} \right) \right],
	\label{eq:indkeld}
  \end{equation}
  for the RK potential, and
  \begin{equation}
	V\left( \sqrt{\rho^2 + D^2} \right) = \frac{k e^2}{\kappa \left( \rho^2 + D^2 \right)}
	\label{eq:indcoul}
  \end{equation}
  for the Coulomb potential.

  Therefore, one can obtain the eigenfunctions and eigenenergies by solving Eq.~\eqref{eq:relschro} using the potential~\eqref{eq:keldyshpot} for direct excitons, or for indirect excitons using either potential~\eqref{eq:indkeld} or~\eqref{eq:indcoul}.

  Both the RK and Coulomb potentials have central symmetry, therefore the eigenstates of the system can be specified by a principal quantum number $n = 1,2,3,\dots$ and an angular momentum quantum number $l = -n+1, -n+2,\dots,0,\dots,n-2,n-1$.
  % For a given principal quantum number $n$, the set of eigenstates with different $l$ are degenerate in the Coulomb system.
  % In the RK system, however, eigenstates with the same $n$ but different $\lvert l \rvert$ are not degenerate \textendash{} the binding energy of the $l=0$ state for a given $n$ is always the smallest, and the binding energies increase as $\lvert l \rvert$ increases, so that, for example, $E_{3d} > E_{3p} > E_{3s}$.
  % This can be justified conceptually by considering the local screening effects of the inhomogeneous dielectric environment which is described by the RK potential.
  % In the case where the dielectric constant of the monolayer(s) in which the excitons are located is larger than the dielectric constant of the environment (in this case we consider the Xenes to be encapsulated by $h$-BN, $\epsilon_{h-\text{BN}}=4.89$), then as the distance between the electron and hole increases, their electrostatic interaction becomes more weakly screened.
  % Since the average electron-hole separation increases as $l$ is increased, for a given $n$ the $l=0$ state is always the most weakly bound, with the states becoming more strongly bound as $l$ increases.
  % Of course, the eigenstates of $(n, \pm l)$ are still two-fold degenerate, but the $n-1$-fold degeneracy which is characteristic of the Coulomb eigensystem is no longer seen in the RK system.
  For the sake of brevity, we shall refer to the eigenstates of the exciton using the familiar nomenclature of the ideal 2D hydrogen atom, that is, $1s$ refers to $(n,l)=(1,0)$, $2p$ would refer to $(n,l) = (2, \pm1)$, and so on.
  This convention is common in the literature~\cite{Cha2016,Berghauser2016,Qiu2013,Poellmann2015b,Wang2015a,Hanbicki2015,Stroucken2015,Zhu2015a,Wu2015b,Steinleitner2017,Berkelbach2013,Brunetti2018}.

  \subsection{\label{ssec:optproptheory}Optical Properties of Excitons in buckled 2D materials}

  Our approach to calculating the optical properties follows well-established methods for modeling optical transitions in atom-like systems~\cite{Snoke}.
  This approach was used to describe the optical absorption by excitons in semiconductor coupled quantum wells~\cite{Lozovik1997}.
  We treat the exciton as a two-level system, modeling its polarization in response to an incident electromagnetic wave as a harmonic oscillator.
  The oscillator strength, $f_0$, of a particular optical transition is given by,
  \begin{equation}
	f_0 = \frac{2 \mu \left( E_{f} - E_{i} \right) \lvert \langle \psi_{f} \vert x \vert \psi_{i} \rangle \rvert^2}{\hbar^2},
	\label{eq:f0}
  \end{equation}
  where $E_i$ and $E_f$ are the eigenenergies corresponding to the eigenfunctions $\psi_i$ and $\psi_f$, and $x$ represents the linear polarization of the electric field of the incident electromagnetic wave.
  The dipole matrix element, $\lvert \langle \psi_f \vert x \vert \psi_i \rangle \rvert^2$, determines which transitions are allowed or forbidden.
  The allowed optical transitions are given by $n_f \neq n_i$ and $l_f = l_i \pm 1$.
  Hence, the allowed transitions from the ground state are those with $n_f = 2,3,\dots$ and $l_f = \pm1$, i.e.\ the states $2p,3p,\dots$.
  The oscillator strength is theoretically useful, as it is a dimensionless quantity which gives the strength of a particular optical transition relative to all other possible transitions from the initial state $\psi_i$.

  Experimentally studying the optical properties of a material generally involves observing how a sample absorbs, transmits, or reflects different wavelengths of electromagnetic waves.
  The intensity of an electromagnetic wave of frequency $\omega$ propagating a distance $z$ through a medium is given by:
  \begin{equation}
	I(z;\omega) = I_0 e^{-\alpha(\omega)z}
	\label{eq:intensity}
  \end{equation}
  where $I_0$ is the original intensity of the wave and $\alpha$ is the absorption coefficient, and is calculated as,
  \begin{equation}
	\alpha(\omega) = \left( \frac{\omega}{\omega_0 c} \frac{\pi e^2}{2 \epsilon_0 \sqrt{\epsilon_{h-\text{BN}}} \mu} \frac{n}{L_X}f_0 \right)\left( \frac{\left( \Gamma/2 \right)}{\left( \omega_0^2 - \omega^2 \right)^2 + \left( \Gamma/2 \right)^2} \right),
	\label{eq:absspec}
  \end{equation}
  where $\omega_0 = (E_f - E_i)/\hbar$ is the Bohr angular frequency of the optical transition, $n$ is the 2D concentration of excitons, $L_X$ represents the thickness of the monolayer(s) which the electron and hole occupy, and $\Gamma$ is the full width half maximum (FWHM) of the optical transition.
  We can deduce from Eq.~\eqref{eq:intensity} that the absorption coefficient is the inverse of the propagation distance $z$ over which the intensity of the electromagnetic wave decreases by a factor $1/e$.

  Evaulating Eq.~\eqref{eq:absspec} for a single optical transition will yield a Lorentzian peak centered on $\omega_0$ with a FWHM of $\Gamma$.
  The absorption spectrum, obtained experimentally by measuring the absorption of a medium across a wide range of incident frequencies $\omega$, is represented theoretically by summing over Eq.~\eqref{eq:absspec} for all possible optical transitions in the medium (not limited to excitonic transitions).
  We focus on the maximal value of the absorption coefficient, obtained when the incident electromagnetic wave is in resonance with a given optical transition, i.e. $\omega = \omega_0$.
  This maximal value is:
  \begin{equation}
	\alpha(\omega = \omega_0) = \left( \frac{\pi e^2}{2 c \epsilon_0 \sqrt{\epsilon_{h-\text{BN}}} \mu} \frac{n}{L_X} f_0 \right) \left( \frac{2}{\Gamma} \right).
	\label{eq:absmax}
  \end{equation}

  However, in 2D materials, where the thickness of a monolayer is a fixed value, the absorption coefficient is not the most efficient way to compare absorption properties across different materials.
  Recalling Eq.~\eqref{eq:intensity}, one can consider the absorption factor, $\mathcal{A} = 1 - I(z=L_X;\omega=\omega_0)/I_0 = 1 - \exp (- \alpha(\omega=\omega_0) L_X)$:
  \begin{equation}
	\mathcal{A} = 1 - \exp \left[ - \left( \frac{\pi e^2 n}{2 c \epsilon_0 \sqrt{\epsilon_{h-\text{BN}}} \mu} f_0 \right) \left( \frac{2}{\Gamma} \right) \right],
	\label{eq:absfac}
  \end{equation}
  which gives the fraction of the electromagnetic wave absorbed by a particular excitonic transition in direct excitons in a single Xene monolayer or in spatially indirect excitons in an X-BN-X heterostructure.

\section{\label{sec:dirres}Direct excitons in Xene Monolayers}

  Below we present the results of calculations for freestanding Xene monolayers as well as monolayer silicene on an $h$-BN substrate.
  The input parameters used in the calculations are given in Table~\ref{tab:matpars}.

  While it is certainly instructive and informative to consider freestanding silicene, germanene, and stanene, it is also important to consider other scenarios which may be experimentally more practical, namely, the behavior of these materials when placed on different substrates.
  Hexagonal boron nitride is a promising substrate for silicene due to its atomically flat structure and relatively weak interactions with the silicene monolayer.
  Indeed, $h$-BN has been identified as an excellent substrate for other 2D materials such as graphene~\cite{Dean2010a,Novoselov2012,Geim2014} and TMDC monolayers~\cite{Thygesen2017,Kou2017,Calman2015}.
  There does, however, appear to be some disagreement regarding exactly how the weak interaction between the $h$-BN and silicene affects the properties of the silicene, if at all.

  The authors of Ref.~\onlinecite{Li2013} performed \textit{ab initio} calculations and found that the interaction between $h$-BN and silicene leads to a rather significant modification of the material properties of silicene, increasing the band gap and decreasing the Fermi velocity of silicene such that its material parameters more closely resemble those of freestanding germanene.
  The authors find that there are nine different stacking arrangements of silicene on $h$-BN, based on the slight lattice mismatch between the two materials, and the variety of different rotation angles between the two lattices.
  All but three of the nine different stacking arrangements result in a bandgap in silicene between $32-39$ meV, and the other three arrangements lead to band gaps of 27, 28, and 29 meV.
  All but one of the lattice arrangements results in a Fermi velocity of at least 92\% of $v_F$ in freestanding silicene, which the authors calculated to be $5.33 \times 10^5~\text{m/s}$.
  One lattice arrangement results in a significantly lower value of the Fermi velocity, only 83\% the magnitude of $v_F$ in freestanding silicene.
  Interestingly, the authors find that the buckling parameter of silicene is not changed by the $h$-BN substrate, but remains constant at $d_0 = 0.46~\text{\AA}$, the same as for freestanding silicene.

  Fortunately, one lattice arrangement has both the largest bandgap and highest Fermi velocity, while a second arrangement has both the smallest band gap and lowest Fermi velocity.
  This allows us to easily provide lower and upper bounds on the calculated properties of excitons in silicene on $h$-BN\@.
  These parameters are presented in Table~\ref{tab:matpars} and are taken from Ref.~\onlinecite{Li2013}.

  Curiously, the authors of Ref.~\onlinecite{Kaloni2013a} also studied silicene on an $h$-BN substrate using \textit{ab-initio} calculations, but arrived at a completely different result compared to Ref.~\onlinecite{Li2013}.
  They find that the buckling parameter of silicene is increased from 0.46 to 0.54 \AA, while they also find that the band gap and Fermi velocity remains largely unchanged compared to freestanding silicene.
  For this reason, we did not perform a separate set of calculations corrsponding to these data, since the results would very closely resemble that of freestanding silicene.

  \subsection{\label{ssec:ebdir}Eigenenergies of direct excitons in monolayer Xenes}

  \begin{figure}[b]
	\centering
	\includegraphics[width=0.49\columnwidth]{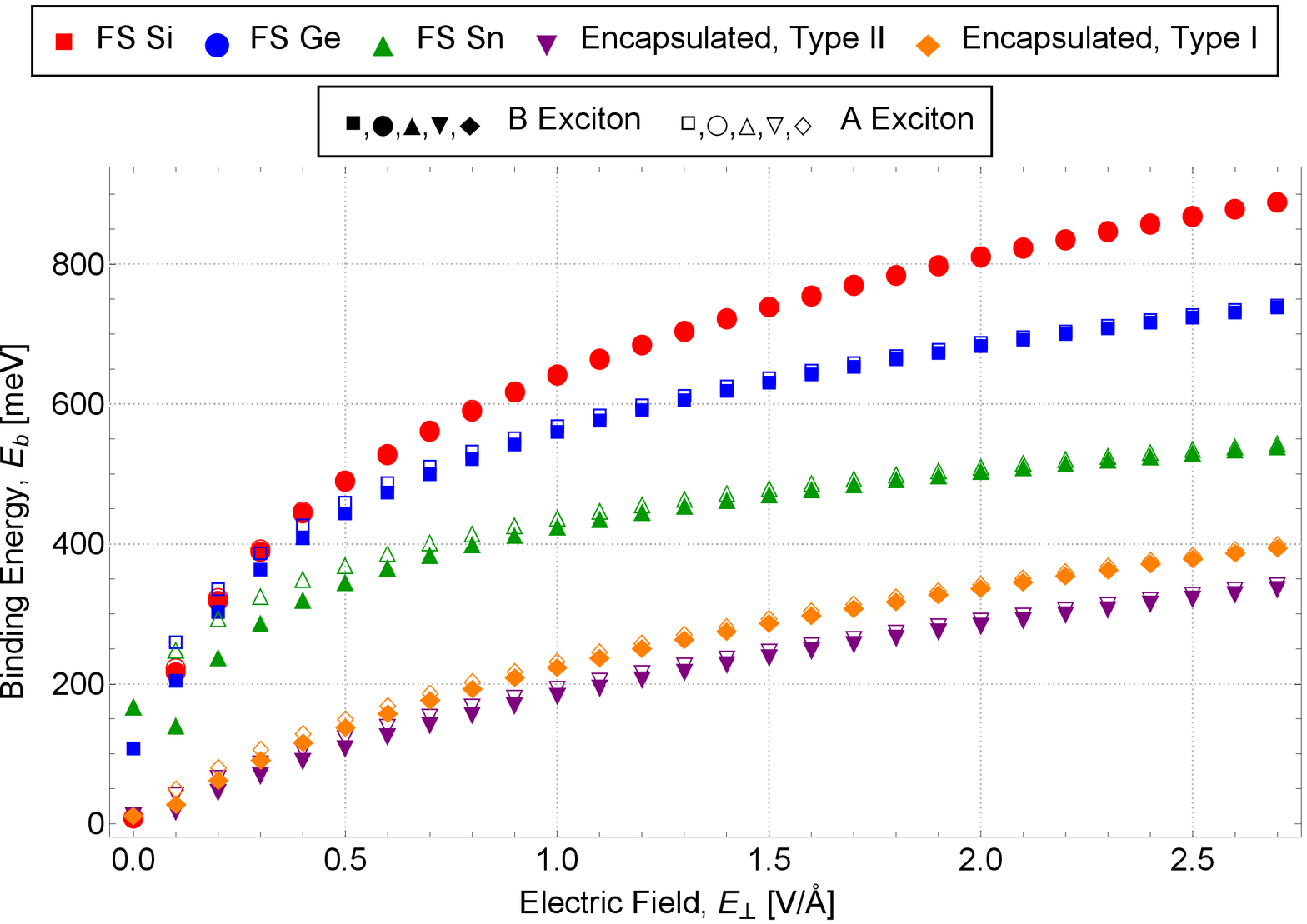}
	\includegraphics[width=0.49\columnwidth]{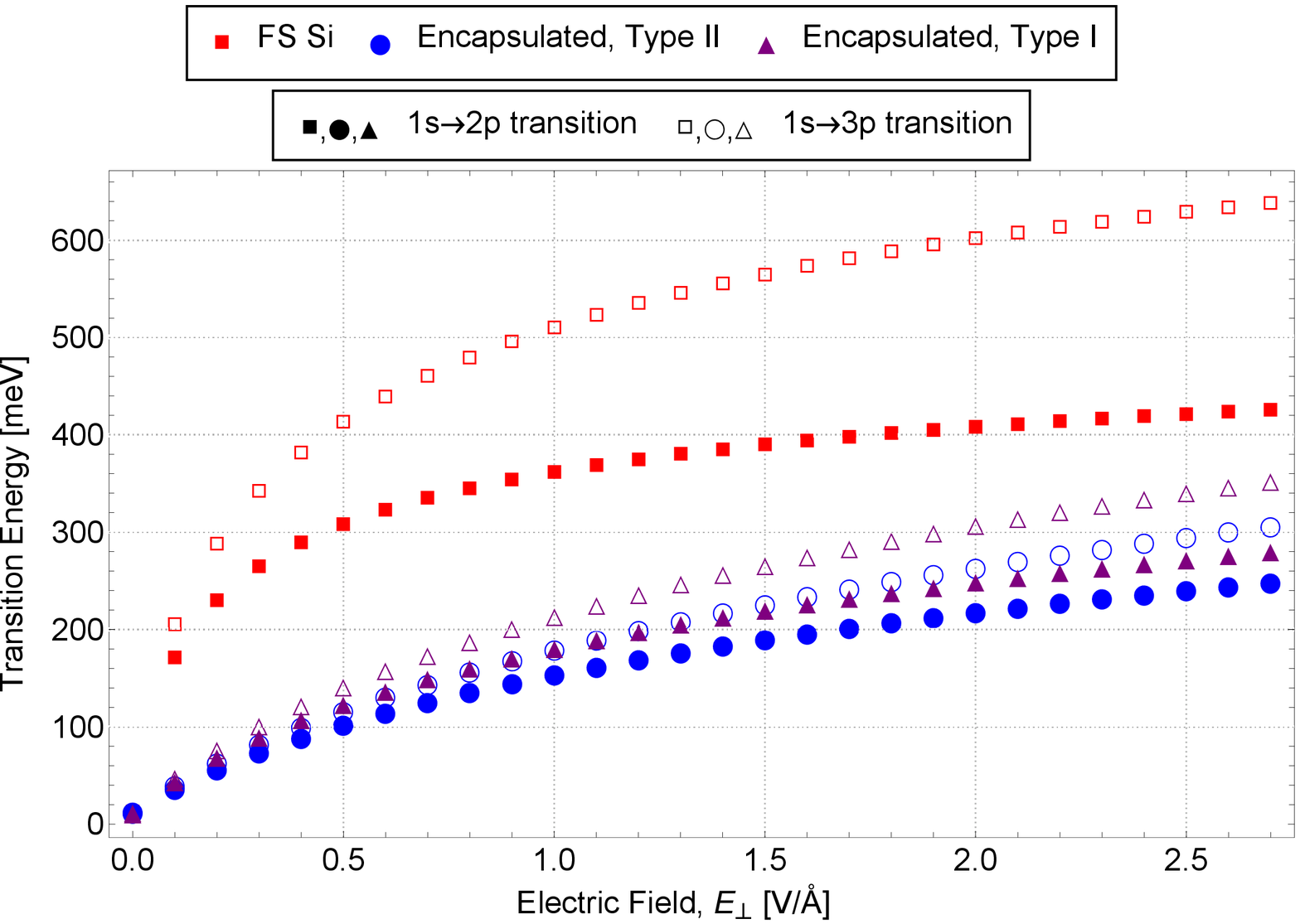}
	\caption{%
	\label{fig:sgs-dir-eb}
	Direct exciton binding energies in freestanding silicene, germanene, and stanene, and in silicene encapsulated by $h$-BN.
	The solid symbols correspond to $B$ excitons, while open symbols represent $A$ excitons.
  }\vspace{-0.3cm}
	\caption{%
	\label{fig:sgs-dir-etr}
	  Dependence of the optical transition energy on the external electric field for the $1s \to 2p$ and $1s \to 3p$ transitions for direct $A$ excitons in freestanding and encapsulated silicene monolayers.
	}
  \end{figure}
  \begin{figure}[t]
	\centering
  \end{figure}

  The results of our calculations for the binding and optical excitation energies of direct excitons in monolayer Xenes are presented in Figs.~\ref{fig:sgs-dir-eb} and~\ref{fig:sgs-dir-etr}, respectively.
  In Fig.~\ref{fig:sgs-dir-eb} we compare the direct exciton binding energy of freestanding silicene, germanene, and stanene, and encapsulated silicene.
  The direct exciton binding energies for FS Ge and FS Sn are qualitatively similar to FS Si, but they are smaller than freestanding Si and larger than encapsulated Si.
  The freestanding monolayers exhibit by far the largest binding energies, due to the much weaker dielectric screening induced by the environment compared to silicene encapsulated by $h$-BN.
  The curves for FS Ge and FS Sn qualitatively resemble that of FS Si, but FS Ge reaches a maximum binding energy of 725 meV, and the maximum binding energy for FS Sn is roughly 525 meV, significantly smaller than for FS Si.
  The percent difference between the binding energy of FS Si and FS Ge at the largest electric field considered, $E_\perp = 2.7~\text{V/\AA}$, is 18.5\%, and the percent difference between FS Si and FS Sn at the same electric field is 47.9\%.
  In addition, in FS Ge and FS Sn, we observe a non-negligible difference in the binding energy of $A$ and $B$ excitons at electric fields up to about $1~\text{V/\AA}$.
  These differences decrease as the electric field increases.
  In FS Si, the difference between $A$ and $B$ excitons is always negligible.

  Overall, we find that FS Si exhibits the largest direct exciton binding energy, followed by FS Ge and then FS Sn, despite the fact that silicene has the lowest-mass charge carriers, while stanene has the highest mass charge carriers.
  Silicene has the lowest mass charge carriers because (a) it has the smallest intrinsic band gap, (b) it has the smallest buckling parameter, so the external electric field induces the smallest change in its band gap, and (c) it has the largest Fermi velocity, again implying that the charge carriers have intrinsically small mass.
  The opposite of points (a), (b), and (c) explain why stanene has the largest carrier masses.
  We infer that stanene has the smallest direct exciton binding energy because it has by far the largest dielectric constant, and the largest monolayer thickness, and therefore, the screening length $\rho_0$ is much larger than in silicene.
  This leads to significant dielectric screening, especially as carrier masses increase and the average exciton radius decreases.
  Direct excitons in $h$-BN-encapsulated silicene show the smallest binding energies, due to the aforementioned strong dielectric screening of the surrounding environment.

  Fig.~\ref{fig:sgs-dir-etr} presents the optical transition energies and reveals an unexpected difference in behavior between freestanding Xenes and encapsulated silicene.
  We use the data for FS Si as representative of the other two FS materials, without making the Figure visually cluttered.
  Since FS Si qualitatively resembles FS Ge and FS Sn, we will only show the results for FS Si throughout the rest of the paper.

  At an electric field greater than $0.6-1.3$ V/\AA, each of the three freestanding materials exhibit a saturation of the $1s \to 2p$ optical transition energy, that is, the transition energy does not change significantly as the electric field continues to increase.
  Furthermore, each of the three FS materials show the same rapid increase at small electric fields.
  In FS Si, we can see that the value of the transition energy at $E_\perp = 1.3~\text{V/\AA}$ is within 10\% of the maximal value, at $E_\perp = 2.7~\text{V/\AA}$.
  In FS Ge, the transition energy approaches 10\% of its maximal value ($E_{max,1s \to 2p} \approx 300~\text{meV}$) at $E_\perp = 1~\text{V/\AA}$.
  In FS Sn ($E_{max,1s \to 2p} \approx 200~\text{meV}$), the same happens at an electric field of only $0.6~\text{V/\AA}$.
  We can infer that this saturation is due to the binding energy and the $2p$ state eigenenergy increasing at roughly the same rate at high electric fields.
  The optical transition energy in the freestanding materials is therefore less tunable than the direct exciton binding energy, since the transition energy for all three FS materials does not change significantly as the electric field is increased.
  In contrast, encapsulated Si continues to show a linear increase in the $1s \to 2p$ optical transition energy even at high electric fields.
  It is also interesting to note that the transition energies of $A$ and $B$ excitons converge to nearly the same value at relatively small electric fields in FS Ge and Sn, even though Fig.~\ref{fig:sgs-dir-eb} demonstrates that the difference in binding energies of $A$ and $B$ excitons in these two materials remains noticable until the electric field becomes larger than the value at which the $A$ and $B$ transition energies converge.
  It would be very interesting to probe these optical transitions experimentally to determine if both the $A$ and $B$ excitonic optical transitions may be induced by a single probe laser tuned to the common transition energy.

  Fig.~\ref{fig:sgs-dir-etr} also shows the $1s \to 3p$ transition energies, which are consistently and significantly larger than the $1s \to 2p$ transition energies.
  Indeed, in all three FS materials, we see that the $1s \to 3p$ transition energy can be up to 50\% larger than the $1s \to 2p$ transition, especially as the electric field approaches its maximum.
  In encapsulated Si, we find that the difference is not so dramatic, but still on the order of 25\% or greater.

  In addition, one can see from Figs.~\ref{fig:sgs-dir-eb} and~\ref{fig:sgs-dir-etr} that the dependence of the eigenenergies of direct excitons calculated using the RK potential on the electric field is non-linear, while the reduced mass of the exciton linearly depends on the electric field according to Eq.~\eqref{eq:effmassEz}.
  It is well known that the eigenenergies of direct 2D excitons calculated with the Coulomb potential are directly proportional to the exciton reduced mass.
  Therefore, in contrast to the RK potential, the eigenenergies of the exciton in the case of the Coulomb potential would depend linearly on the electric field.

  \subsection{\label{ssec:optdir}Optical properties of direct excitons}

\begin{figure}[t]
	\centering
	\includegraphics[width=0.49\columnwidth]{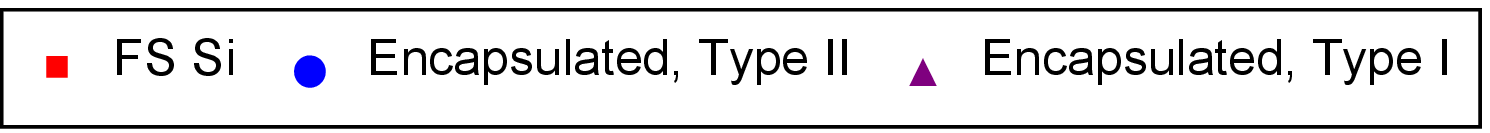}
	\includegraphics[width=0.43\columnwidth]{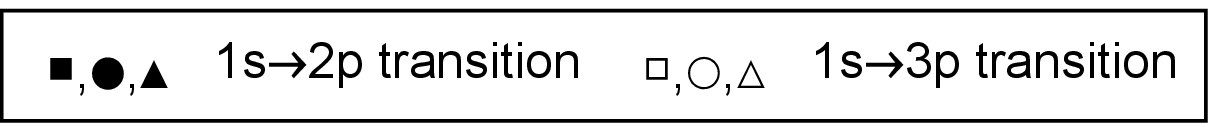}
	\includegraphics[width=0.32\columnwidth]{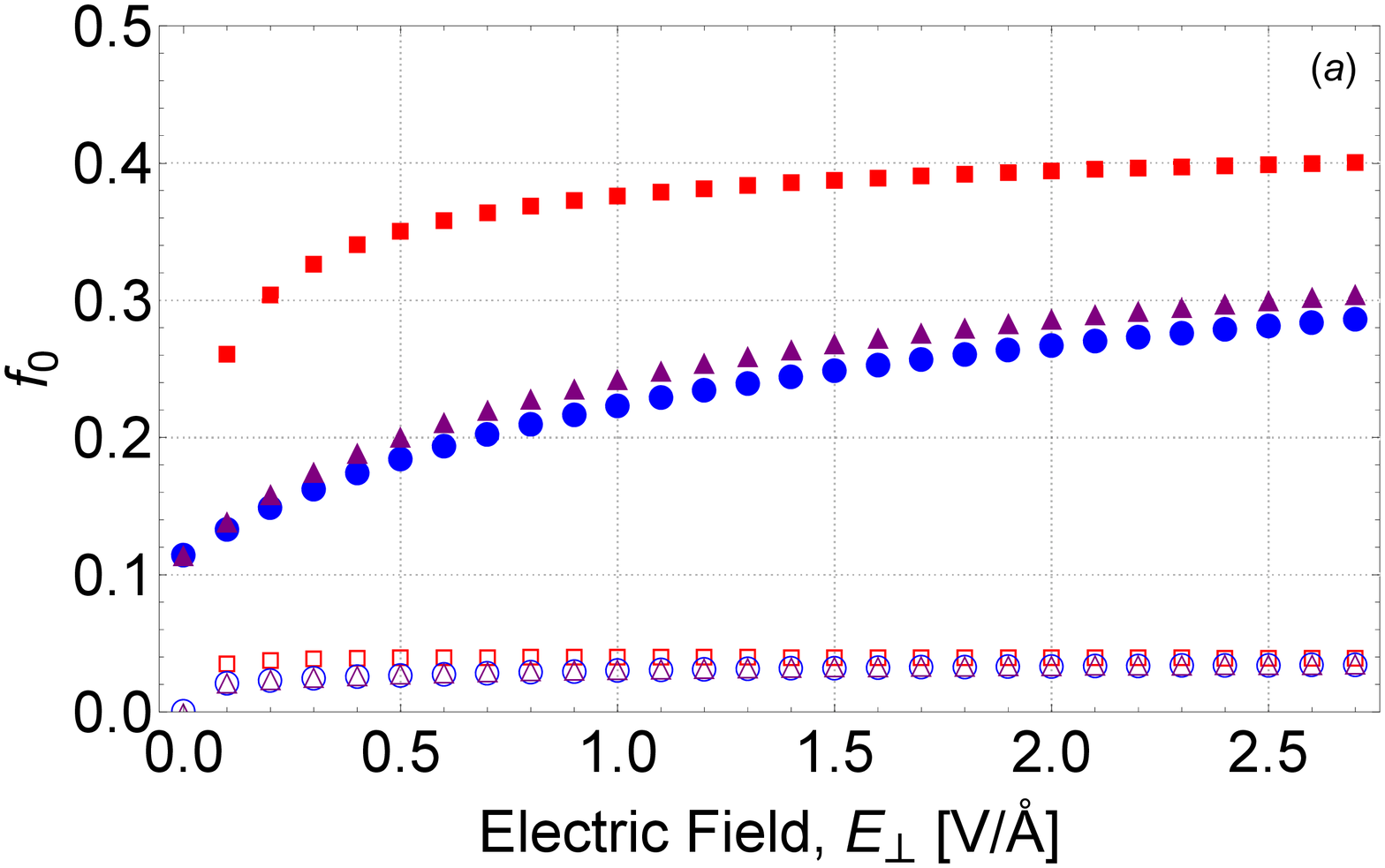}
	\includegraphics[width=0.335\columnwidth]{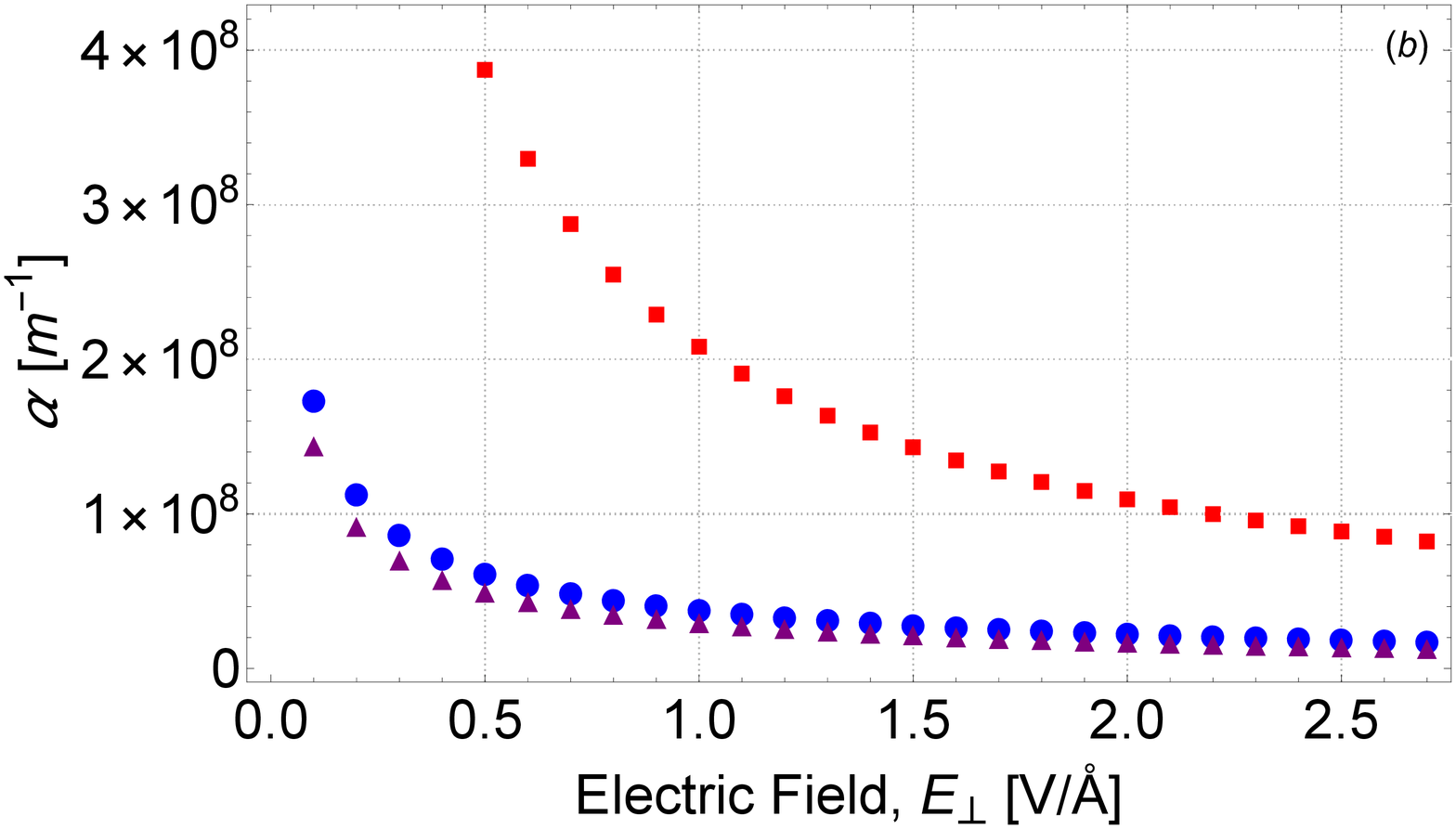}
	\includegraphics[width=0.32\columnwidth]{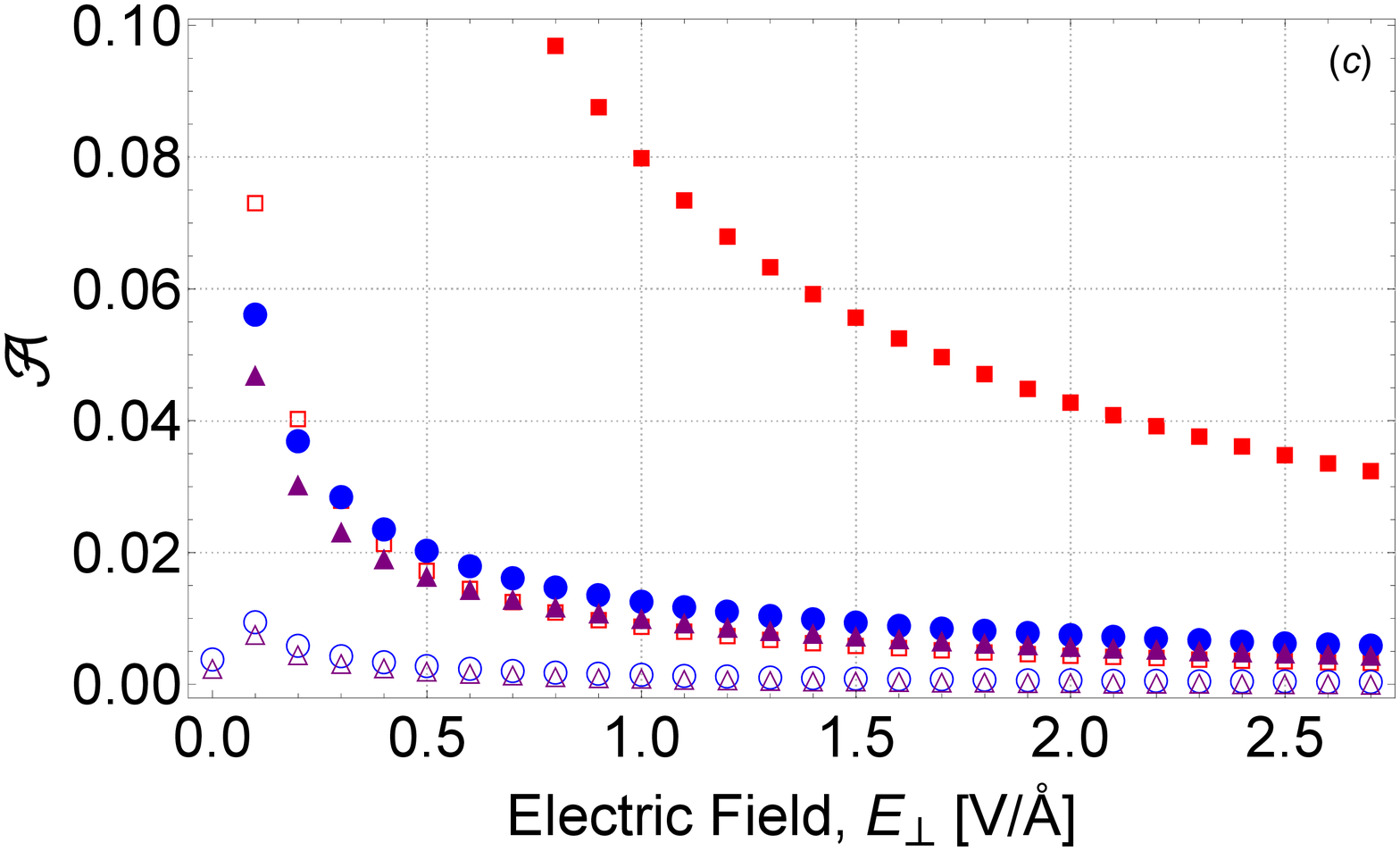}
	\caption{%
	  Optical properties of direct excitons in Xene monolayers.
	  The dependence of the (a) the oscillator strength, $f_0$, (b) the absorption coefficient, $\alpha$, and (c) the absorption factor, $\mathcal{A}$, on the electric field, $E_{\perp}$, for direct $A$ excitons in FS and encapsulated Si.
	In (a) and(c) both the $1s \to 2p$ and $1s \to 3p$ transitions are shown, while in (b), only the $1s \to 2p$ optical transition is shown.}
	\label{fig:sgs-dir-absafac}
  \end{figure}

  The results of calculations of the optical properties of direct excitons in monolayer Xenes are presented in Fig.~\ref{fig:sgs-dir-absafac}.
  The oscillator strengths of the three freestanding materials quickly become saturated at a value of about $0.4$, as shown in Fig.~\ref{fig:sgs-dir-absafac}a.
  Furthermore, there is very little difference in $f_0$ for a given material for $A$ and $B$ excitons.
  The oscillator strengths in encapsulated Si never quite reach saturation, and never come close to the same magnitude as that of the freestanding materials.

  The oscillator strengths for the $1s \to 3p$ optical transition are also given in Fig.~\ref{fig:sgs-dir-absafac}a.
  Surprisingly, the behavior of $f_{1s \to 3p}$ as a function of the external electric field is qualitatively very similar to $f_{1s \to 2p}$.
  We find that the value of $f_{1s \to 3p}$ is roughly one-tenth the magnitude of the corresponding value of $f_{1s \to 2p}$ at a given electric field.
  This consistent difference in magnitude of roughly a factor of 10 is somewhat surprising, considering the rather small magnitudes of $f_{1s \to 2p}$ at electric fields less than approximately 1 V/\AA.
  It was thought that perhaps this would mean that $f_{1s \to 3p}$ would be of comparable magnitude to $f_{1s \to 2p}$ at small electric fields, however this is clearly not the case.

  On the other hand, in FS Si, $f_{1s \to 3p}$ quickly reaches a value of 0.04 at small electric fields, but we observe a very slight decrease in the magnitude of $f_{1s \to 3p}$ as the electric field continues to increase beyond approximately 1 V/\AA\@.
  This slight decrease in $f_0$ at electric fields greater than $\approx 1$ V/\AA~resembles the observed behavior of $f_{1s \to 2p}$ for indirect excitons in X-BN-X heterostructures of FSE Sn, when the interlayer separation is large and the electric field is strong.

  The absorption coefficient and absorption factor for FS and encapsulated Si are shown in Figs.~\ref{fig:sgs-dir-absafac}b and~\ref{fig:sgs-dir-absafac}c, respectively.
  We observe that the freestanding materials should absorb significantly more light than encapsulated Si.
  This again can likely be tied back to the difference in dielectric environment \textendash{} recall from Eq.~\eqref{eq:absmax} the factor of $\sqrt{\epsilon_{h-\text{BN}}}$ in the denominator.
  For freestanding Xenes in a vacuum, this would translate to significantly stronger absorption than for $h$-BN encapsulated materials.
  It is also noteworthy that the absorption in encapsulated silicene becomes saturated by the electric field much more quickly, not exhibiting much change when the electric field is increased beyond $E_\perp = 1~\text{V/\AA}$.
  On the other hand, FS Si exhbits a noticable change in its absorption through the entire range of electric fields.
  In FS Ge, $\alpha$ and $\mathcal{A}$ lies roughly between the curves for FS Si and encapsulated Si.
  In FS Sn, $\alpha$ converges towards encapsulated Si at large electric fields, while $\mathcal{A}$ remains larger than in encapsulated Si by approximately one percentage point.

  Ultimately, we find that the minimum absorption factor, obtained at the maximum value of the electric field, is approximately 3\% for FS Si, 2\% for FS Ge, 1.5\% for FS Sn, and only about 1\% for encapsulated silicene.
  For the sake of comparison, at an electric field of $E_\perp = 1~\text{V/\AA}$, the value of the absorption factor in FS Ge is approximately 5.5\% (where $B$ excitons absorb slightly more, and $A$ excitons absorb slightly less), and in FS Sn, the corresponding value is slightly less than 4\% for $B$ excitons, but slightly more than 3\% for $A$ excitons.
  As the electric field is increased, the difference between $A$ and $B$ excitons becomes much less significant.
  The FS materials show a stronger response in their optical absorption as a function of electric field, suggesting that they are more tunable than encapsulated Si, which approaches its global minimum at an electric field of about 1.5 V/\AA.

  Surprisingly, the $1s \to 3p$ transition in freestanding Si is comparable to the $1s \to 2p$ transition in encapsulated Si.
  On the other hand, the $1s \to 3p$ transition in encapsulated Si is quite strongly suppressed, barely surpassing 1\% absorption at an electric field of on 0.1 V/\AA, and decreasing to only a small fraction of 1\% absorption as the electric field continues to increase.

  \section{\label{sec:ind}Properties of Indirect Excitons in X-BN-X heterostructures}

  In the following subsections, we study the dependence of the binding energy and optical properties of spatially indirect excitons on the external electric field, $E_\perp$, as well as on the number of $h$-BN monolayers in the X-BN-X heterostructure.
  We continue to perform calculations using the parameters corresponding to freestanding Si, Ge, and Sn, even though it is of course unreasonable to expect the Xene monolayers to retain their freestanding parameters when placed in an X-BN-X heterostructure.
  In the following calculations, we now consider the dielectric environment $\kappa = 4.89$, unlike the case of direct excitons, where the truly freestanding materials were modeled to be surrounded by vacuum, i.e. $\kappa=1$.
  To clearly denote the difference between calculations for direct excitons in freestanding monolayers, and calculations using the freestanding parameters in an X-BN-X heterostructure, we will refer to the latter as freestanding-encapsulated, or FSE.
  We shall present our results for indirect excitons in FSE materials in an X-BN-X heterostructure as a means of illustrating the importance of using physically accurate material parameters when calculating the properties of indirect excitons.

  \subsection{\label{ssec:optpropind}Eigenenergies of spatially indirect excitons in an X-BN-X heterostructure}

\begin{figure}[h]
	\begin{center}
	  \includegraphics[width=0.45\columnwidth]{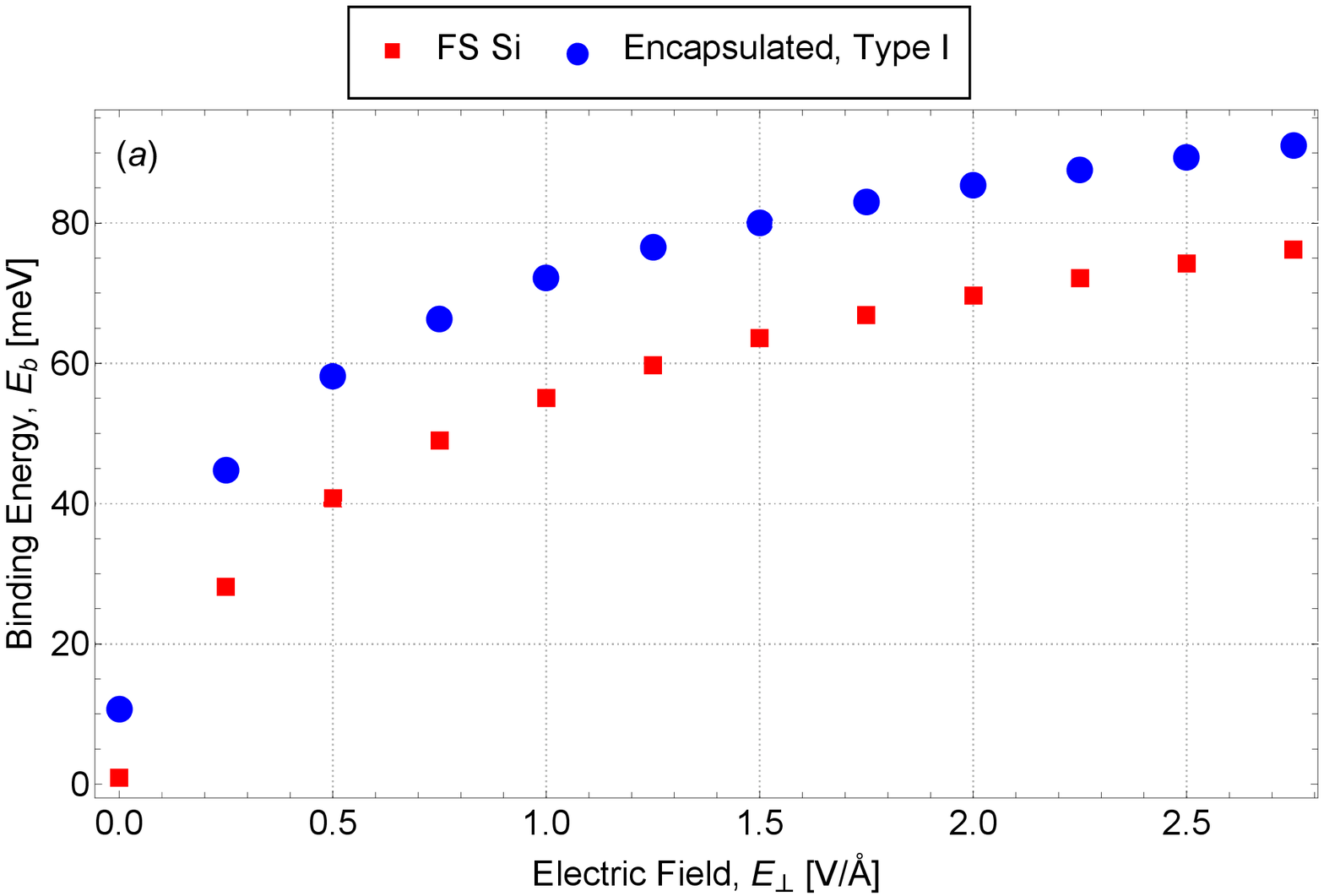}
	  \includegraphics[width=0.45\columnwidth]{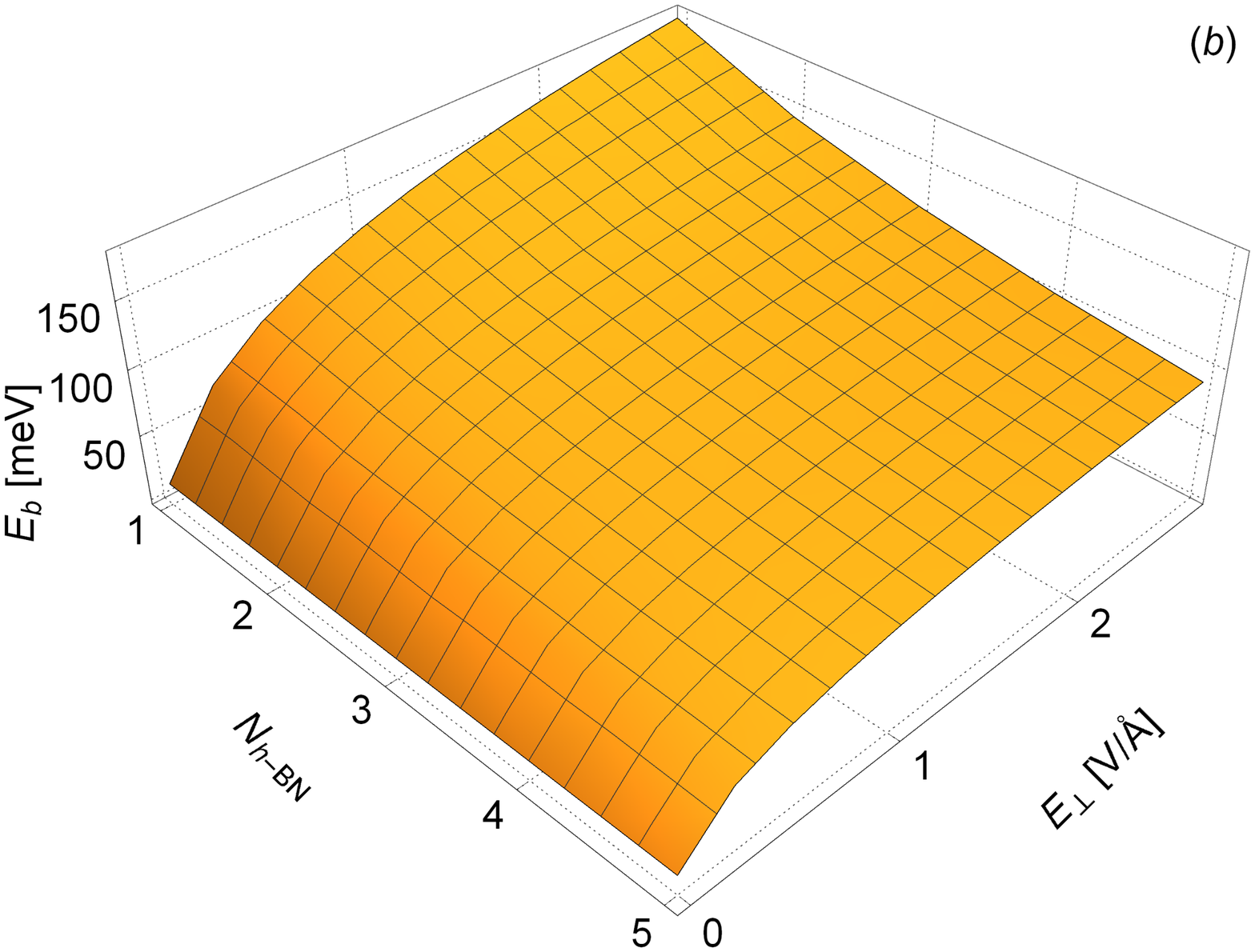}
	\end{center}\vspace{-0.4cm}
	\caption{%
	  (a) Indirect $A$ exciton binding energy as a function of external electric field at $N=5$ for FSE and encapsulated Si.
	  (b) Dependence of the indirect $A$ exciton binding energy in encapsulated Si on the interlayer separation, $N$ and the external electric field, $E_\perp$.
	  Calculations are performed using the Rytova-Keldysh potential.
	}
	\label{fig:sgs-ind-eb}
  \end{figure}

  Fig.~\ref{fig:sgs-ind-eb}a shows the binding energies of indirect $A$ excitons in FSE and encapsulated Si.
  Therefore, the larger intrinsic band gap and significantly smaller Fermi velocity of the encapsulated Si in turn leads to consistently larger binding energies than the FSE Si at all values of electric field and interlayer separation.
  Even at large interlayer separation, we see that there is a significant difference in the binding energy between FSE and encapsulated Si, and this difference between the binding energies increases significantly as the interlayer separation decreases.
  Therefore, the observed difference in indirect exciton binding energy in Fig.~\ref{fig:sgs-ind-eb}a of greater than 10\% at $N=5$ is even more pronounced at smaller interlayer separations.
  In Fig.~\ref{fig:sgs-ind-eb}b, it is shown that the binding energy of indirect $A$ excitons in encapsulated Si increases sharply as the electric field is increased up to about 1 V/\AA, but as the electric field continues to increase, the binding energy does not increase significantly.
  Increasing the interlayer separation from $N = 1$ to $N=5$ reduces the binding energy by about 33\% at high electric fields.

\begin{figure}[h]
	\centering
	\includegraphics[width=0.49\columnwidth]{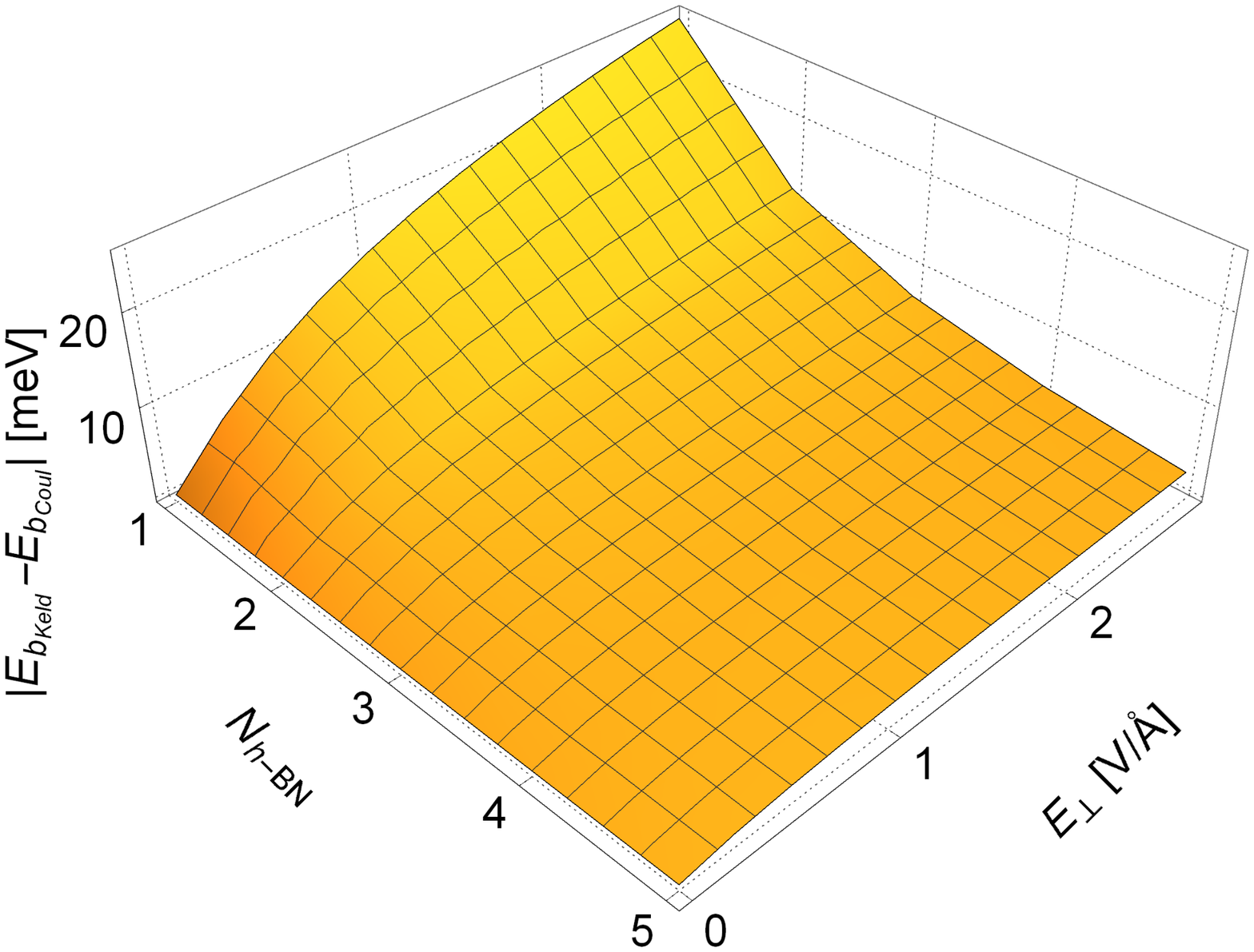}
	\includegraphics[width=0.49\columnwidth]{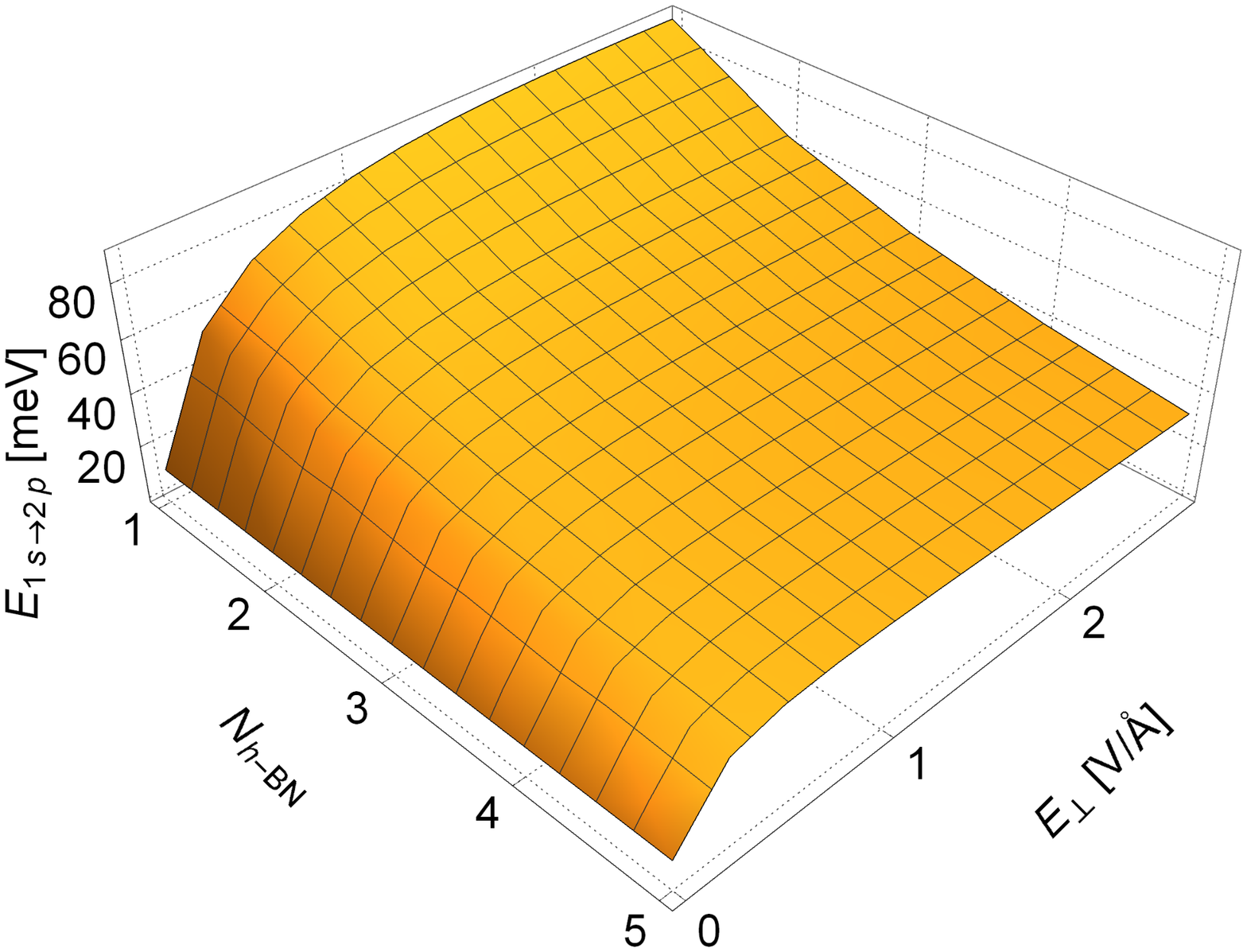}
	\caption{%
	\label{fig:sgs-ind-eb-ckcompare}
	  Difference in binding energy for indirect $A$ excitons calculated using Coulomb and Rytova-Keldysh potentials in encapsulated Type~\rn{2} Si, as a function of the interlayer distance, $N$, and the external electric field, $E_\perp$.
	}\vspace{-0.3cm}
	\caption{%
	\label{fig:sgs-ind-etr}
	  Dependence of the $1s \to 2p$ optical transition energies of indirect $A$ excitons in encapsulated Type II Si on the interlayer distance, $N$, and external electric field, $E_\perp$.
	  Calculations are performed using the RK potential.
	}
  \end{figure}

  To understand the role of screening we perform calculations for the Coulomb and RK potentials.
  Fig.~\ref{fig:sgs-ind-eb-ckcompare} provides a comparison of the value of the binding energy in Type II encapsulated Si using either the Coulomb or RK potentials.
  We find that the binding energies calculated with the Coulomb potential are always larger than those calculated using the RK potential.
  For one monolayer of $h$-BN, the percent difference in the binding energies for FSE Si range from roughly 5\% at small applied electric fields (5.35\% at $E_\perp = 0.25~\text{V/\AA}$) up to nearly 12\% at the maximum calculated electric field (11.7\% at $E_\perp = 2.75~\text{V/\AA}$).
  For the same values of the electric field and interlayer separation, the percent difference in the binding energies for FSE Ge is more prominent than in FSE Si, starting at 10.9\% at $E_\perp = 0.25~\text{V/\AA}$, and increasing up to 20.1\% at $E_\perp = 2.75~\text{V/\AA}$.
  The percent difference in FSE Sn is by far the most pronounced, beginning at 19.6\% and increasing to 34.5\% as the external electric field is increased.

  As one might expect, these differences in the binding energy decrease as the interlayer separation increases.
  This is due to the fact that the RK potential converges towards the Coulomb potential at large distances.
  For an interlayer separation of $N=5$ in FSE Ge, the percent difference ranges between 3.7\% and 5.8\%.
  In FSE Sn, however, the percent difference ranges from 7.2\% to 12.5\%, which is still rather significant.

  The $1s \to 2p$ optical transition energies of indirect $A$ excitons in Type II encapsulated Si are presented in Fig.~\ref{fig:sgs-ind-etr}.
  Curiously, at large $N$, the optical transition energy in encapsulated silicene begins to decrease slightly as the electric field continues to increase.
  Furthermore, our calculations show again that the difference in the optical transition energy between the Coulomb and RK potentials is again quite significant, which reinforces the importance of using a physically accurate interaction potential when calculating the properties of the indirect exciton eigensystem.
  Finally we see that the optical transition energy in encapsulated Si is not, in fact, particularly tunable, since the transition energy plateaus at low electric field at all values of the interlayer separation, and it remains mostly constant, though it does decrease slightly for large interlayer separations and strong electric fields.

  In constrast to the results in Fig.~\ref{fig:sgs-ind-eb}, our calculations show that the difference in the optical transition energy between FSE and encapsulated Si is quite small, on the order of 5\%, even at small interlayer separations and large electric fields.
  Indeed, the optical transition energies calculated using the Coulomb potential at large electric fields and at $N=1$ are remarkably similar for each of the three FSE materials as well as encapsulated Si, with all quantities falling between 90-110 meV.
  On the other hand, when calculated using the RK potential, the transition energy at $N=1$ and at the maximal electric field shows a larger variation between the FS Xenes and encapsulated Si, falling between 60 meV for FSE Sn, 85 meV for FSE Si, and 90 meV for encapsulated Si.
  This result suggests that the optical transition energy is not as sensitive to the choice of material parameters as the indirect exciton binding energy.

	\subsection{\label{ssec:indopt}Optical properties of indirect excitons}

\begin{figure*}[h]
	\centering
	\includegraphics[width=0.32\columnwidth]{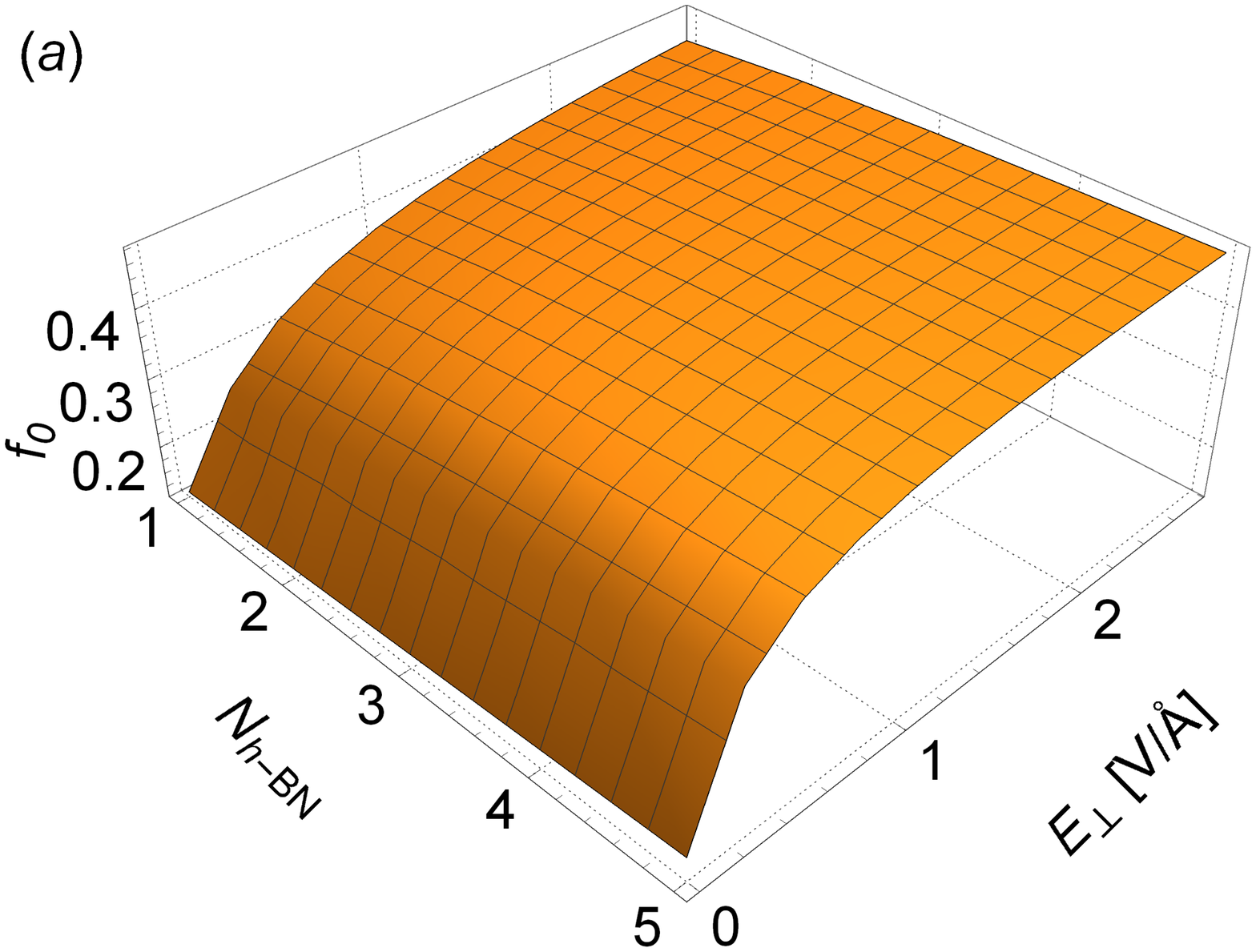}
	\includegraphics[width=0.32\columnwidth]{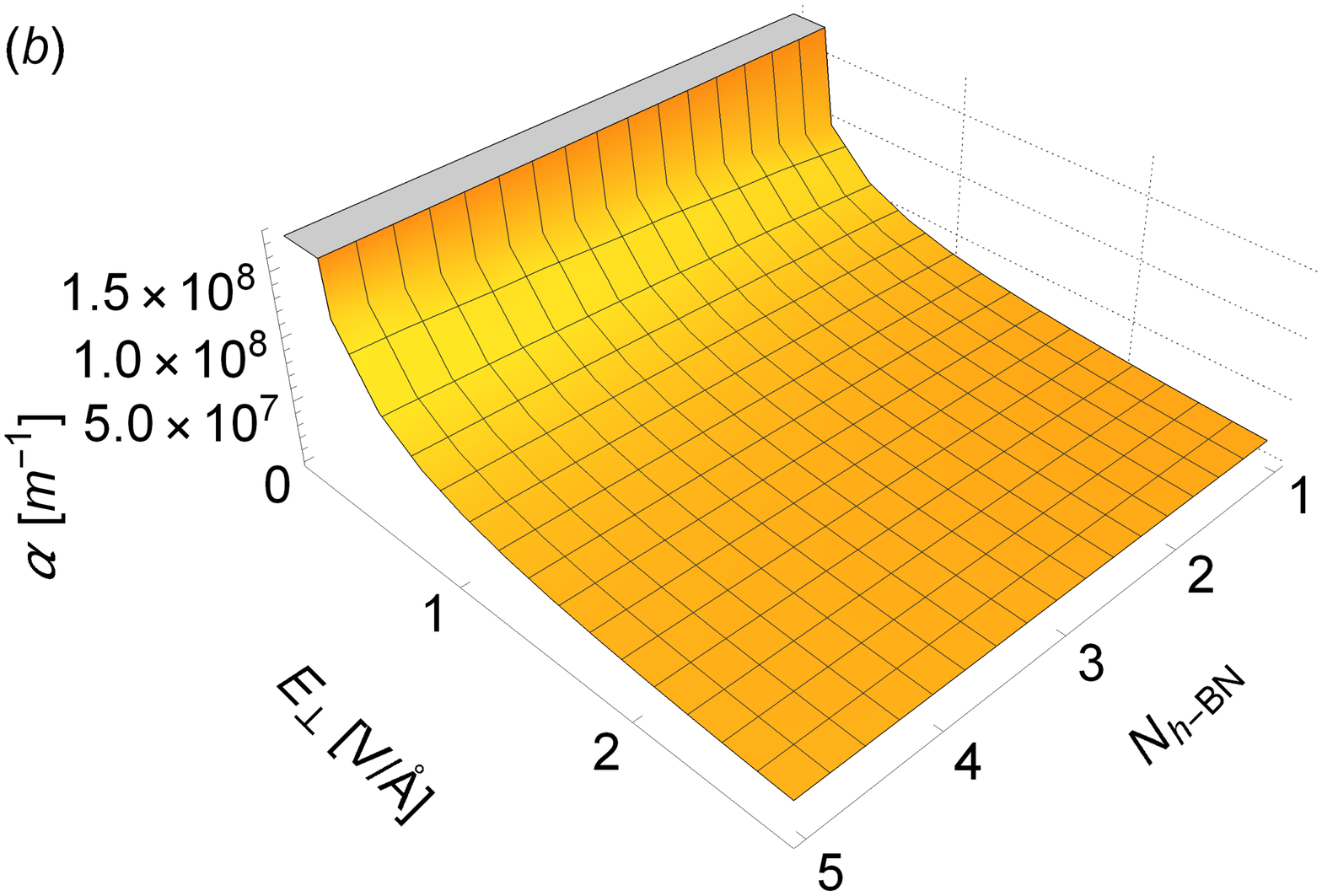}
	\includegraphics[width=0.32\columnwidth]{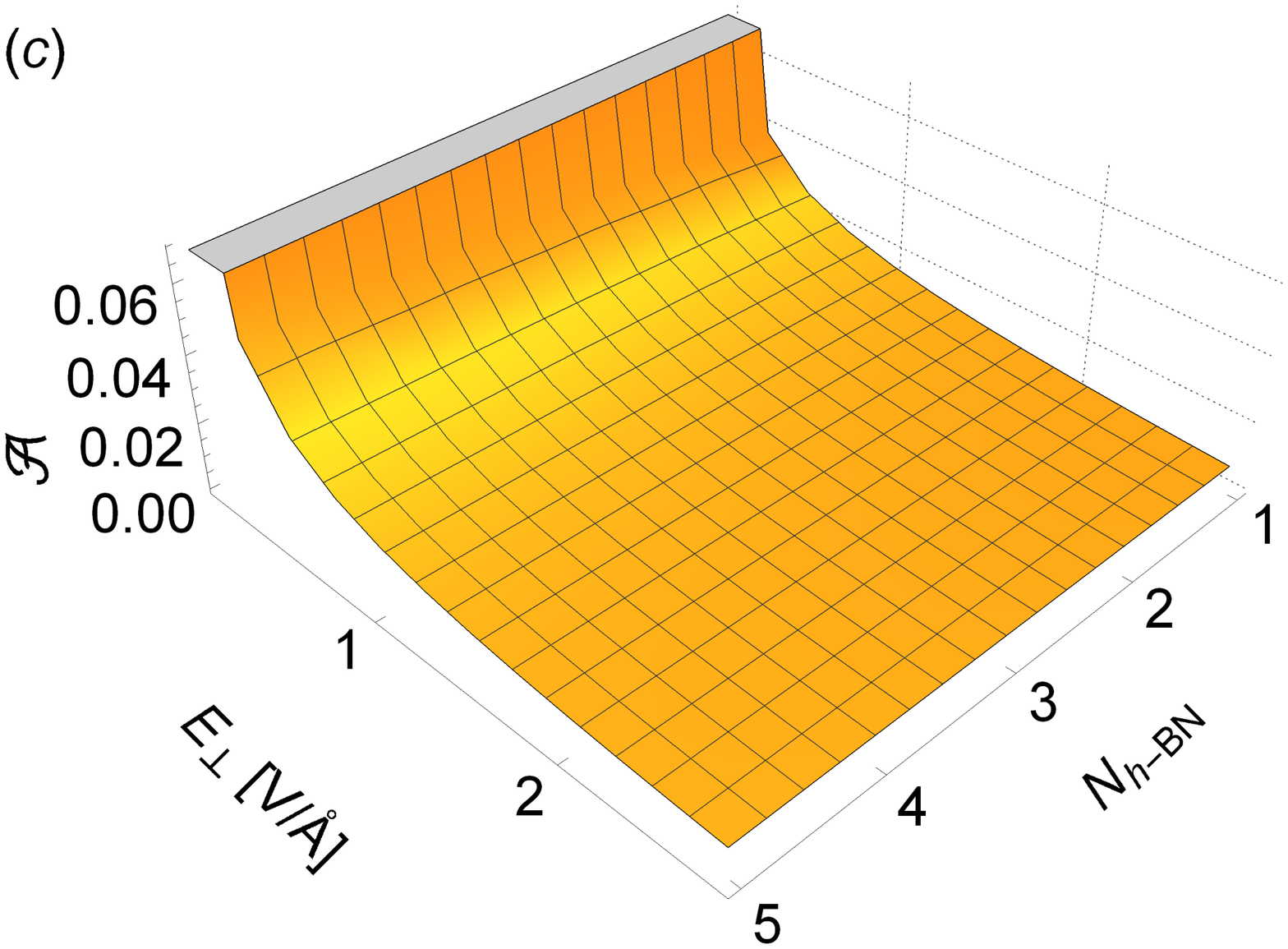}
	\caption{%
	  Optical properties of indirect excitons in X-BN-X heterostructures.
	  The dependence of (a) the oscillator strength, $f_0$, (b) the absorption coefficient, $\alpha$, and (c) the absorption factor, $\mathcal{A}$, for indirect $A$ excitons in encapsulated silicene, on the interlayer separation, $N$, and the external electric field, $E_\perp$.
	  Calculations are performed using the RK potential.
	}
  \label{fig:sgs-ind-absafac}
\end{figure*}

	The results of calculations of the optical properties of indirect excitons in X-BN-X heterostructures are presented in Fig.~\ref{fig:sgs-ind-absafac}.
	The oscillator strength $f_0$ of the $1s \to 2p$ optical transition of spatially indirect $A$ excitons in encapsulated Si increases monotonically with both $E_\perp$ and $D$, as shown in Fig.~\ref{fig:sgs-ind-absafac}a.
  It was previously reported~\cite{Brunetti2018} that $f_0$ is expected to increase monotonically with $N$ for spatially indirect excitons in TMDC-BN-TMDC heterostructures, and this phenomenon is observed for encapsulated Si, as well as for the three FSE materials.
  We find that the oscillator strength approaches 0.5 at large electric fields and interlayer separations, suggesting that the $1s \to 3p$ transition is very strongly suppressed in this regime.
  $f_0$ increases quickly for small values of $E_\perp$ and grows much more slowly beyond around $E_\perp = 1~V/\text{\AA}$.
  At large $N$, $f_0$ quickly approaches $0.5$, which implies that the $1s \to 3p$ optical transition is very strongly suppressed.

  The three FSE materials, not shown in Fig.~\ref{fig:sgs-ind-absafac}, are quantitatively very similar to encapsulated Si.
  This is another example of a quantity which is mostly insensitive to the choice of material parameters used in the calculations.

  Unlike the dramatic difference between the Coulomb and RK potentials seen in the eigenenergies of Fig.~\ref{fig:sgs-ind-eb-ckcompare}, the difference in $f_0$ between the Coulomb and RK potentials is quite small.
  In general, while there is some variation using these potentials between the materials studied here, the quantitative difference is very slight overall, except in FSE Si, where there is still a noticable difference even as the electric field approaches its maximum.

  The oscillator strengths of the $1s \to 3p$ transitions in encapsulated Si were also calculated.
  We find that $f_{1s \to 3p}$ is approximately one-tenth the magnitude of the corresponding $f_{1s \to 2p}$ for a given electric field and interlayer separation, very similar to the case of direct excitons in Xene monolayers.

  Also noteworthy is the unusual, and unique, behavior of $f_0$ at zero electric field for the four materials.
  In FSE Si and Ge, the oscillator strength of the $1s \to 2p$ transition can exceed 0.5, an unphysical result which would appear to violate the oscillator strength sum rule.
  In FSE Si, with its very small intrinsic gap and very large Fermi velocity, we sometimes observe unusual results at very small electric fields, such as the unreasonably large oscillator strength observed here.
  This may be due to the extremely small exciton mass at these small fields, which in turn leads to a huge excitonic radius, which then may run into problems with our computational framework, specifically the size of our computational ``box''.

	% \begin{figure}[b]
	%   \centering
	%   \includegraphics[width=0.60\columnwidth]{614-ind-f0-3d.eps}
	%   \caption{%
	%   Dependence of the oscillator strength for the $1s \to 2p$ optical transition of indirect $A$ excitons in encapsulated Type II Si, on the interlayer separation, $N$, and the external electric field, $E_\perp$, calculated using the RK potential.
	% }
	%   \label{fig:sgs-ind-f0}
	% \end{figure}

  Figs.~\ref{fig:sgs-ind-absafac}b and~\ref{fig:sgs-ind-absafac}c demonstrate how optical absorption is suppressed by increasing the external electric field, just as is the case with direct excitons, as shown in Figs.~\ref{fig:sgs-dir-absafac}b and~\ref{fig:sgs-dir-absafac}c.
  As the interlayer separation is increased, the absorption increases by a small amount.
  These calculations for encapsulated silicene are quantitatively very similar to the three FSE materials.
  Our calculations show that at large electric fields, encapsulated Si, as well as the three FSE materials, should absorb less than 2\% of incoming resonant light.
  Encapsulated Si shows much weaker absorption than FSE Si, with encapsulated Si more closely resembling FSE Ge in terms of its absorption properties.
  As is the case with the other optical quantities, we find that the absorption for encapsulated Si decreases sharply for $E_\perp \lesssim 1$ V/\AA, and more slowly for $E_\perp > 1$ V/\AA~compared to the FSE materials.

  We also see that changing the interlayer separation can have quite a significant effect at low electric fields, but has almost no effect at large electric fields.
  For example, at $E_\perp = 0.5~\text{V/\AA}$, we calculate that the absorption factor in encapsulated Si increases from about 2.5\% at $N = 1$ to about 3.5\% at $N=5$.

  Our calculations also show that there is a small difference in the absorption factor when comparing the Coulomb and RK potentials.
  This difference is on the order of a few tenths of a percent at $E_\perp \lesssim 1$ V/\AA, and becomes negligible as the electric field increases beyond this point.

  Similarly, we find that the difference in absorption between $A$ and $B$ excitons can be quite large at small electric fields, and furthermore than $B$ excitons always absorb more strongly than $A$ excitons, due to their slightly smaller mass.
  Finally, we note that the reduction in $f_0$ by a factor of 10 carries over to the calculated values of $\alpha_{1s \to 3p}$ and $\mathcal{A}_{1s \to 3p}$, as well.

  \section{\label{sec:dirindcomp}Comparison between direct and indirect excitons}

  The binding energies of direct excitons are, of course, stronger than the binding energies of the spatially indirect excitons in the same materials, but this drop in binding energy when moving from direct excitons to indirect excitons is huge.
  For example, the direct exciton binding energy in FS Si is on the order of 900 meV at $E_{\perp} = 2.7$ V/\AA, while the indirect exciton binding energy at $N=1$ in FSE Si is only 140 meV (155 meV) for the RK (Coulomb) potential, a staggering reduction in the binding energy of over 80\%.
  This change is not as drastic in FS Ge and FS Sn, where the binding energy drops by slightly less than 80\% in FS Ge (from $\approx 700$ meV to $\approx 160$ meV) and by about 75\% in FS Sn (from $\approx 550$ meV to $\approx 170$ meV).
  The dramatic difference between direct and indirect binding energies in the freestanding materials is due to both the change in dielectric environment as well as the increase in the electron-hole distance.
  The difference between direct and indirect exciton binding energies is not as severe in encapsulated Si, where the change in binding energy is only about 50\% at the maximum electric field.
  The significantly smaller change in encapsulated Si can be partially explained by $\kappa$ remaining constant at 4.89 between the direct and indirect exciton cases.

  The difference in the optical transition energy for direct and indirect excitons is not as large as the aforementioned difference between the binding energies.
  In FS Si, we see a drop in the transition energy of about 75\% when moving from direct excitons to indirect excitons at $N=1$.
  In FS Ge, the same change is approximately 66\%, and in FS Sn, the difference is only about 50\%.
  Unlike the binding energies, where encapsulated silicene exhibited the smallest direct-to-indirect change, when comparing the transition energies we find that encapsulated silicene shows a change of about 66\%, comparable to FS Ge.

  The oscillator strengths follow a consistent pattern, with the direct excitons having the smallest $f_0$ at any given $E_\perp$, and $f_0$ increasing as the interlayer distance is increased.
  As mentioned in Sec.~\ref{ssec:optproptheory}, we observe that while the oscillator strength increases monotonically with $E_\perp$, increasing quickly at first before slowly leveling off above $E_\perp \gtrsim 1~\text{V/\AA}$, the absorption coefficient displays nearly the opposite behavior.
  For both the direct and indirect exciton cases, we find that the absorption coefficient $\alpha$ decreases monotonically with $E_\perp$ but still increases monotonically with the interlayer separation.
  The same is true of the absorption factor, $\mathcal{A}$.

  On the other hand, the optical properties behave in the opposite way compared to the eigenenergies with respect to the electric field.
  At small electric fields, the difference in, for example, the oscillator strength can be significant, on the order of 10\% or more.
  As the electric field is increased, this difference decreases, and the magnitudes coverge towards each other.
  The absorption coefficient and absorption factor exhibit this same behavior, but these differences can be traced directly back to how the oscillator strength changes, since there are no other terms in the analytical forms of $\alpha$ and $\mathcal{A}$ which would change depending on the choice of interaction potential.
  Due to the oscillator sum rule, we know that the maximum value of the oscillator strength for a given symmetric, photon-absorbing transition must be 0.5.
  Therefore, as the electric field increases, the oscillator strength must approach 0.5, regardless of the interaction potential used.

  Regarding the choice of the RK or Coulomb potentials, we find huge differences in the binding and optical transition energies for interlayer separations $N < 2$, but this difference decreases sharply beyond $N = 3$.
  This significant difference at small interlayer separations is clearly due to the way in which the two potentials treat the surrounding dielectric environment.
  When using the Coulomb potential, the dielectric constant is effectively $\epsilon_{h-\text{BN}} = 4.89$, while the RK potential still takes into account the screening length of the Xene monolayers.
  Since the Xenes have much larger dielectric constants than the $h$-BN, using the RK potential for indirect excitons results in much smaller binding energies when compared to the Coulomb potential.

  Analyzing the relationship between $f_0$ and $E_\perp$ in the context of Coulomb and RK potentials is not as straightforward as our analysis of the eigenenergies.
  This is because $f_0$ is directly proportional to both the transition energy and the dipole matrix element, both of which depend directly on the choice of interaction potential.
  Ultimately, we observe that $f_0$ calculated with the RK potential is always larger than $f_0$ calculated using the Coulomb potential at small electric fields \textendash{} therefore, despite the fact that the optical transition energy is always larger for the Coulomb potential than for the RK potential, it must be the case that the dipole matrix element integral is always much larger for the RK potential than for the Coulomb potential.

  This difference in behavior \textendash{} where the difference between RK and Coulomb increases in the eigenenergies as the electric field increases, while the  difference in the optical properties decreases as the electric field increases \textendash{} suggests a complicated relationship between the choice of interaction potential and the external electric field.
  With regards to the differences in the eigenenergies, we can understand why that difference increases as the electric field increases if we recall that the exciton radius is proportional to the excitonic reduced mass.
  At small electric fields, the exciton has a small mass and therefore a large excitonic radius.
  At large separations, the RK potential converges towards the Coulomb potential, and therefore the difference in eigenenergies calculated using the two potentials is small at small electric fields.
  As the electric field increases, the excitonic reduced mass increases, which reduces the exciton radius, which in turn causes the  eigenenergies calculated using the RK and Coulomb potentials to diverge from each other.

  We note that there is significant disagreement in the literature as to the exact value of the material parameters for FS Xenes given in Table~\ref{tab:matpars}.
  For example, the intrinsic band gap of silicene has been reported to be in the range $1.55-7.9$ meV~\cite{Tabert2014,Liu2011,Liu2011a,Drummond2012}, the germanene band gap has been cited as between $24-93$ meV~\cite{Tabert2014,Liu2011,Liu2011a}, and the band gap in stanene has been reported in Ref.~\onlinecite{Chen2016} to be between $30-123$ meV.
  At large electric fields, these huge discrepancies in the band gap would have a minor effect on the eigenenergies and optical properties of both direct and spatially indirect excitons in silicene, while the differences in germanene and stanene are noticable but minimal.
  At small electric fields, however, these differences in the intrinsic band gap can completely change the type of behavior one would expect to observe.

  The Fermi velocity of charge carriers in Xene monolayers also shows significant variation between results.
  For example, \textit{ab initio} calculations performed in Ref.~\onlinecite{Bechstedt2012} found that in FS Si, $v_F = 5.32 \times 10^{5}~\text{m/s}$, while in FS Ge, $v_F = 5.17 \times 10^{5}~\text{m/s}$.
  These values are considerably smaller than the parameters given in Table~\ref{tab:matpars} and used in our calculations, though these $v_F$ are comparable in magnitude to $v_F$ in encapsulated Si~\cite{Li2013}.
  The significant difference in these values of $v_F$ has a major effect on the charge carrier mass \textendash{} while the two values of $v_F$ in FS Si only differ by about 20\%, the carrier masses in FS Si calculated with $v_F = 5.32 \times 10^5~\text{m/s}$ are 49\% larger than the carrier masses calculated with $v_F = 6.5 \times 10^5~\text{m/s}$.
  This difference of nearly 50\% in the carrier masses would therefore noticeably increase the exciton binding energy while decreasing the absorption.

  Finally, data on the Xene monolayer thickness is scarce, and the data that does exist can vary wildly in magnitude.
  For example, experimental measurements of Si monolayer thickness on various substrates using atomic force microscopy (AFM) yield thicknesses of 0.3 nm~\cite{Grazianetti2016}, 0.37 nm~\cite{Oughaddou2015}, and 0.4 nm~\cite{Derivaz2015}.
  It seems reasonable to expect that a freestanding germanene monolayer would be thicker than a freestanding Si monolayer, since Ge has a larger atomic radius, $R_{\text{Ge}}=1.25~\text{\AA}$~\cite{Slater1964,Slater1965}, than silicene, $R_{\text{Si}}=1.11~\text{\AA}$~\cite{Slater1964,Slater1965}, and germanene has a larger buckling constant by about 0.2~\AA.
  Likewise, freestanding stanene should similarly be thicker than freestanding germanene by roughly the same amount, again because it has a larger atomic radius, $R_{\text{Sn}}=1.45~\text{\AA}$~\cite{Slater1964,Slater1965}, and larger buckling constant, again by about 0.2~\AA.
  Using $l_{\text{Si}} = 0.4~\text{nm}$~\cite{Tao2015} as a baseline, we then arrive at rough estimates of the monolayer thicknesses of FS Ge and FS Sn of $0.45$ nm and $0.5$ nm, respectively.
  Overestimating the monolayer thickness would have the effect of reducing the exciton binding energy but increasing the absorption coefficient.

  We therefore made an effort to choose material parameters for the freestanding Xenes which were provided from a single source in order to maintain the consistency of our results.

  \section{\label{sec:conc}Conclusions}

  In this paper we demonstrate that an external electric field can be used to tune the eigenenergies and optical properties of direct and indirect excitons in Xene monolayers or X-BN-X heterostructures.
  Reflecting upon our results, we see that this is generally true, with the condition that most quantities in the FS Xenes reach a saturation point at some value of the electric field, beyond which the value of the quantity does not change by much as the electric field continues to increase.
  Specifically, we find that in the freestanding Xenes, the optical transition energies and oscillator strengths saturate at low electric fields, while in encapsulated Si, it is the absorption coefficient and absorption factor that become saturated at low electric fields.
  For indirect excitons in X-BN-X heterostructures, we observe saturation of the oscillator strengths, absorption coefficients, and absorption factors.

  In addition, our study of indirect excitons using both the Coulomb and RK potentials to describe the electrostatic interaction of the electron and hole has indicated that the choice of interaction potential can cause huge changes in the magnitude of the binding energies and optical transition energies, making it imperative that theorists determine which interaction potential yields physically accurate results.
  The eigenenergies calculated using the Coulomb potential are always larger than the corresponding quantities calculated using the RK potential, and this difference increases as the electric field increases.
  Conversely, the optical properties calculated using the RK potential are always of higher magnitude than the corresponding values calculated using the Coulomb potential, though this difference is negligible at large electric fields.

  Finally, our comparison of the properties of indirect excitons calculated using the material parameters of freestanding Si and using the material properties of Si with $h$-BN as a substrate show that the choice of material parameters does indeed have a significant effect on the eigensystem, and that it would therefore be physically inaccurate to treat the Xene parameters as unchanged between the freestanding monolayer and an X-BN-X heterostructure.

  These calculations provide a reference for future theoretical and experimental studies of intraexcitonic optical transitions.
  In addition, our calculations demonstrate that further studies are necessary to expand and refine our understanding of the tunability of excitons in 2D Xenes.
  The comparison of the exciton properties in FSE Si and encapsulated Si demonstrate that it is necessary to correctly identify the material parameters of the Xenes, in particular the band gap, Fermi velocity, and effective monolayer thickness.
  It is especially important to examine how these properties change when the Xene monolayer is placed on different substrates, and how, if at all, these parameters change as a function of the external electric field.
  The difference in the eigenenergies and optical properties of indirect excitons calculated used the Coulomb and RK potentials provides an opportunity for further study of the role of screening effects.
  These interesting topics will need to be explored further, as they may play an important role in the use of 2D Xenes in novel nanodevices.

\bibliography{sgspublish.bib}

\end{document}